\documentclass[MSNbibl,dvips]{arxstspdf}
\usepackage{flushend}
\usepackage{stfloats}
\usepackage{graphicx}
\volume{26}
\issue{1}
\pubyear{2011}
\firstpage{130}
\lastpage{149}
\doi{10.1214/11-STS354}

\begin{document}
\begin{frontmatter}

\title{Variable Selection for Nonparametric Gaussian Process Priors: Models and Computational Strategies}
\runtitle{Variable Selection for Gaussian Process Models}

\begin{aug}
\author[a]{\fnms{Terrance} \snm{Savitsky}\ead[label=e1]{tds151@gmail.com}},
\author[b]{\fnms{Marina} \snm{Vannucci}\ead[label=e2]{marina@rice.edu}}
\and
\author[c]{\fnms{Naijun} \snm{Sha}\ead[label=e3]{nsha@utep.edu}}

\runauthor{T. Savitsky, M. Vannucci and N. Sha}

\affiliation{RAND Corporation, Rice University, University of Texas at El Paso}

\address[a]{Terrance Savitsky is Associate Statistician, RAND Corporation, 1776 Main Street, Santa Monica, California 90401-3208, USA \printead{e1}.}
\address[b]{Marina Vannucci is Professor, Department of Statistics, Rice University, 6100 Main Street, Houston, Texas 77030, USA \printead{e2}.}
\address[c]{Naijun Sha is Associate Professor, Department of Mathematical Sciences, University of Texas at El Paso, 500 W University Ave, El Paso, Texas 79968, USA \printead{e3}.}
\end{aug}

\begin{abstract}
This paper presents a unified treatment of Gaussian process models that
extends to data from the exponential dispersion family and to survival
data. Our specific interest is in the analysis of data sets with
predictors that have an a priori unknown form of possibly nonlinear
associations to the response. The modeling approach we describe
incorporates Gaussian processes in a generalized linear model framework
to obtain a class of nonparametric regression models where the
covariance matrix depends on the predictors. We consider, in
particular, continuous, categorical and count responses. We also look
into models that account for survival outcomes. We explore alternative
covariance formulations for the Gaussian process prior and demonstrate
the flexibility of the construction. Next, we focus on the important
problem of selecting variables from the set of possible predictors and
describe a general framework that employs mixture priors. We compare
alternative MCMC strategies for posterior inference and achieve a
computationally efficient and practical approach. We demonstrate
performances on simulated and benchmark data sets.
\end{abstract}


\end{frontmatter}

\section{Introduction}

In this paper we present a unified modeling approach to Gaussian
processes (GP) that extends to data from the exponential dispersion
family and to survival data.  With the advent of kernel-based methods,
models utilizing Gaussian processes have become very common in machine
learning approaches to regression and classification problems; see
\citet{rasm2006}. In the statistical literature GP regression models
have been used as a~nonparametric approach to model the nonlinear
relationship between a response variable and a set of predictors; see,
for example, \citet{ohagan1978}.  \citet{sacks2000} employed a
stationary GP function of spatial locations in a regression model to
account for residual spatial variation.  \citet{diggle1998} extended
this construction to model the link function of the generalized linear
model (GLM) construction of \citet{mccullagh1989}. \citet{neal1999}
considered linear regression and logit models.

We follow up on the literature cited above and introduce Gaussian
process models as a class that broadens the generalized linear
construction by incorporating fairly complex continuous response
surfaces. The key idea of the construction is to introduce latent
variables on which a Gaussian process prior is imposed. In the general
case the GP construction replaces the linear relationship in the link
function of a GLM.  This results in a class of nonparametric regression
models that can accommodate linear and nonlinear terms, as well as
noise terms that account for unexplained sources of variation in the
data.  The approach extends to latent regression models used for
continuous, categorical and count data. Here we also consider a class
of models that account for survival outcomes. We explore alternative
covariance formulations for the GP prior and demonstrate the
flexibility of the construction. In addition, we address practical
computational issues that arise in the application of Gaussian
processes due to numerical instability in the calculation of the
covariance matrix.

Next, we look at the important problem of selecting variables from a
set of possible predictors and describe a general framework that
employs mixture priors. Bayesian variable selection has been a topic of
much attention among researchers over the last few years. When a large
number of predictors is available the inclusion of noninformative
variables in the analysis may degrade the prediction results. Bayesian
variable selection methods that use mixture priors were  investigated
for the linear regression model by \citeauthor{george1993} (\citeyear{george1993,george1997}), with
contributions by various other authors on special features of the
selection priors and on computational aspects of the method; see
\citet{autokey1} for a nice review. Extensions to linear
regression models with multivariate responses were put forward by
\citet{brown1998} and to multinomial probit by \citet{sha2004}. Early
approaches to Bayesian variable selection for generalized linear models
can be found in \citet{chen1999} and \citet{raftery1995}. Survival
models were considered by \citet{volinsky1997} and, more recently, by
\citet{lee2004} and \citet{sha2006}.  As for Gaussian process models,
\citet{link2006} investigated Bayesian variable selection methods in
the linear regression framework by employing mixture priors with a
spike at zero on the parameters of the covariance matrix of the
Guassian process prior.

Our unified treatment of Gaussian process models extends the line of
work of \citet{link2006} to more complex data structures and models. We
transform the covariance parameters and explore designs and MCMC
strategies that aim at producing a minimally correlated parameter space
and efficiently convergent sampling schemes. In particular, we find
that Metropolis-within-Gibbs schemes achieve a substantial improvement
in computational efficiency. Our results on simulated data and
benchmark data sets show that GP models can lead to improved
predictions without the requirement of pre-specifying higher order and
nonlinear additive functions of the predictors. We show, in particular,
that a Gaussian process covariance matrix with a single exponential
term is able to map a mixture of linear and nonlinear associations with
excellent prediction performance.

GP models can be considered part of the broad class of nonparametric
regression models of the type $y=f(\mathbf{x})+\mathit{error}$, with $y$ an
observed (or latent) response, $f$ an unknown function and $\mathbf{x}$ a
$p$-dimensio\-nal vector of covariates, and where the objective is to
estimate the function $f$ for prediction of future responses. Among
possible alternative choices to GP models, one famous class is that of
kernel regression models, where the estimate of $f$ is selected from
the set of functions contained in the reproducing kernel Hilbert space
(RKHS) induced by a chosen kernel. Kernel models have a long and
successful history in statistics and machine learning [see
\citet{parzen1963}, \citet{wahba1990} and \citet{cristianini2004}] and
include many of the most widely used statistical methods for
nonparametric estimation, including spline models and methods that use
regularized techniques. Gaussian processes can be constructed with
kernel convolutions and, therefore, GP models can be seen as contained
in the class of nonparametric kernel regression with exponential family
observations. \citet{rasm2006}, in particular, note that the GP
construction is equivalent to a linear basis regression employing an
infinite set of Gaussian basis functions and results in a response
surface that lies within the space of all mathematically smooth, that
is, infinitely mean square differentiable, functions spanning the RKHS.
Constructions of Bayesian kernel methods in the context of GP models
can be found in \citet{bishop2006} and \citet{rasm2006}.

Another popular class of nonparametric spline regression models is the
generalized additive models (GAM) of \citet{ruppert2003}, that employ
linear projections of the unknown function $f$ onto a set of basis
functions, typically cubic splines or B-splines, and related
extensions, such as the structured additive regression (STAR) models
of\break
\citet{fahrmeir2004} that, in addition, include interaction surfaces,
spatial effects and random effects. Generally speaking, these
regression mo\-dels impose additional structure on the predictors and are
therefore better suited for the purpose of interpretability, while
Gaussian process models are better suited for prediction. Extensions of
STAR models also enable variable selection based on spike and slab type
priors; see, for example, \citet{panagiotelis2008}.

Ensamble learning models, such as bagging, boosting and random forest
models, utilize decision trees as basis functions; see
\citet{hastie2001}. Trees readily model interactions and nonlinearity
subject to a maximum tree depth constraint to prevent overfitting.
Generalized boosting models (GBMs), as an example, such as the Ada\-Boost
of \citet{freund1997}, represent a~nonlinear function of the covariates
by simpler basis~func\-tions typically estimated in a stage-wise,
iterative fashion that successively adds the basis functions to fit
generalized or pseudo residuals obtained by mini\-mizing a chosen loss
function. GBMs accommodate dichotomous, continuous, event time and
count res\-ponses.  These models would be expected to produce similar
prediction results to GP regression and classification models. We
explore their behavior on one of the benchmark data sets in the
application section of this paper. Notice that GBMs do not incorporate
an explicit variable selection mechanism that allows to exclude
nuisance covariates, as we do with GP models, although they do provide
a relative measure of variable importance, averaged over all
trees.

Regression trees partition the predictor space and fit independent
models in different parts of the~\mbox{input} space, therefore facilitating
nonstationarity and leading to smaller local covariance matrices.
``Treed GP'' models are constructed by \citet{gramacy2008} and extend
the constant and linear construction of \citet{chipman2002}. A~prior is
specified over the tree process, and posterior inference is performed
on the joint tree and leaf models. The effect of this formulation is to
allow the correlation structure to vary over the input space. Since
each tree region is composed of a portion of the observations, there is
a computational savings to generate the GP covariance matrix from
$m_{r} < n$ observations for region $r$.  The authors note that treed
GP models are best suited ``$\ldots$towards problems with a smaller
number of distinct partitions$\ldots.$''  So, while it is theoretically
possible to perform variable selection in a forward selection manner,
in applications these models are often used with single covariates.

The rest of the paper is organized as follows: In Section~\ref{ggpmod}
we formally introduce the class of GP models by broadening the
generalized linear construction. We also extend this class to include
models for survival data. Possible constructions of the GP covariance
matrix are enumerated in Section~\ref{GPcov}. Prior distributions for
variable selection are discussed in Section~\ref{priors} and posterior
inference, including MCMC algorithms and prediction strategies, in
Section~\ref{posteriors}.  We include simulated data illustrations for
continuous, count and survival data regression in Section~\ref{Simulation}, followed by benchmark applications in Section~\ref{applications}. Concluding remarks and suggestions for future
research are in Section~\ref{discussion}. Some details on computational
issues and related pseudo-code are given in the
\hyperref[app]{Appendix}.

\vspace*{6pt}
\section{Gaussian Process Models}\label{ggpmod}
\vspace*{6pt}

We introduce Gaussian process models via a unified modeling approach
that extends to data from the exponential dispersion family and to
survival data.

\vspace*{2pt}
\subsection{Generalized Models}
\vspace*{2pt}

In a generalized linear model the monotone link function $g(\cdot)$
relates the linear predictors to the cano\-nical parameter as
$g(\eta_i)=\mathbf{x}_{i}'{\bolds \beta}$, with $\eta_{i}$ the canonical
parameter for the $i$th observation, $\mathbf{x}_i=(x_1,\ldots,x_p)'$ a
$p\times 1$ column vector of predictors for the $i$th subject and
${\bolds\beta}$ the coefficient vector ${\bolds
\beta}=(\beta_1,\ldots,\beta_p)'$. A~broader class of models that
incorporate fairly complex continuous response surfaces is obtained by
introducing latent variables on which a Gaussian process prior is
imposed. More specifically, the latent variables $z(\mathbf{x}_i)$ define
the values of the link function as
\begin{equation}\label{ggpm}
g(\eta_i)=z(\mathbf{x}_i), \quad i=1,\ldots,n,
\end{equation}
and a Gaussian process (GP) prior on the $n\times 1$ latent vector is
specified as
\begin{equation}\label{ZvecDist}
\mathbf{z}(\mathbf{X})=(z(\mathbf{x}_1),\ldots, z(\mathbf{x}_n))' \sim N( \mathbf{0},\mathbf{C}),
\end{equation}
with the $n\times n$ covariance matrix $\mathbf{C}$ a fairly
complex
function of the predictors. This class of models can be cast within the
model-based geostatistics framework of \citet{diggle1998}, with the
dimension of the space being equal to the number of covariates.

The class of models introduced above extends to latent regression
models used for continuous, categorical and count data. We provide some
details on models for continuous and binary responses and for count
data, since we will be using these cases in our simulation studies
presented below. GP regression models are obtained by choosing the link
function in (\ref{ggpm}) as the identity function, that is,
\begin{equation}\label{uni}
\mathbf{y} =  \mathbf{z}(\mathbf{X}) + \bolds{\varepsilon},
\end{equation}
with $\mathbf{y}$ the $n\times 1$ observed response vector,
$\mathbf{z}(\mathbf{X})$ an $n$-dimensional realization from a GP as in
(\ref{ZvecDist}), and $\bolds{\varepsilon} \sim \mathcal{N}(0,
\frac{1}{r}\mathbb{I}_{n})$ with $r$ a precision parameter. A~Gamma
prior can be imposed on $r$, that is, $r \sim
\mathcal{G}(a_{r},b_{r})$. Linear models of type (\ref{uni}) were
studied by \citet{neal1999} and \citet{link2006}. One notices that, by
integrating $\mathbf{z}(\mathbf{X})$ out, the marginalized likelihood
is
\begin{equation}\label{unidist}
\mathop{\mathbf{y}|\mathbf{C},r} \sim \mathcal{N}\biggl(\mathbf{0},
\biggl[\frac{1}{r}\mathbb{I}_{n}+\mathbf{C}\biggr]\biggr),
\end{equation}
that is, a regression model with the covariance matrix of the response
depending on the predictors.~Non\-linear response surfaces can be
generated as a~function of those covariates for suitable choices of the
covariance matrix. We discuss some of the most popu\-lar in Section~\ref{GPcov}.

In the case of a binary response, class labels $t_{i} \in \{0,1\}$ for
$i=1,\ldots,n$ are observed. We assume $t_{i}
\sim\operatorname{Binomial}(1;p_{i})$ and define
$p_i=P(t_{i}=1|z(\mathbf{x}_{i}))$ with
$\mathbf{z}(\mathbf{X})$ as in (\ref{ZvecDist}). For logit models, for
example, we have $p_{i}=\mathbf{F}(z(\mathbf{x}_{i}))=1/[1 +
\exp(-z(\mathbf{x}_{i}))]$. Similarly, for binary probit we
can directly define the inverse link function as $p_{i} =
\Phi(z(\mathbf{x}_{i}))$, with $\Phi(\cdot)$ the cdf of
standard normal distribution. However, a more common approach to
inference in probit models uses data augmentation; see
\citet{albert1993}. This approach defines latent values $y_{i}$ which
are related to the response via a regression model, that is, in our
latent GP framework, $y_{i} = z(\mathbf{x}_{i})+ \varepsilon_{i}$, with
$\varepsilon_{i} \sim \mathcal{N}(0,1)$, and associated to the observed
classes, $t_{i}$, via the rule $t_i=1$ if $y_{i} > 0$ and $t_i=0$ if
$y_{i} < 0$. Notice that the latent variable approach results in a~GP
on $\mathbf{y}$ with a covariance function obtained by adding a~``jitter''
of variance one to $\mathbf{C}$, with a similar effect of the noise
component in the regression models (\ref{uni}) and (\ref{unidist}).
\citet{neal1999} argues that an effect close to a~probit model can be
produced by a logit model~by introducing a large amount of jitter in
its covariance matrix. Extensions to multivariate models for continuous
and categorical responses are quite straightforward.

As another example, count data models can be obtained by choosing the
canonical link function for the Poisson distribution as
$\log(\bolds{\lambda}) =\mathop{\mathbf{z(\mathbf{X})}}$ with
$\mathbf{z(\mathbf{X})}$ as in~(\ref{ZvecDist}). Over-dispersion,
possibly caused from lack of inclusion of all possible predictors, is
taken into account by modeling the extra variability via random
effects, $u_{i}$, that is, $\tilde{\lambda}_{i} = \exp(z(\mathbf{x}_{i}) +
u_{i})=\break\exp(z(\mathbf{x}_{i}))\exp(u_{i}) = \lambda_{i}\delta_{i}$.  For
identifiability, one can impose $\mathbb{E}(\delta_{i})=1$ and
marginalize over $\delta_{i}$ using a~conjugate prior, $\delta_{i} \sim
\mathcal{G}(\tau,\tau)$, to achieve the negative binomial likelihood as
in \citet{long1997},
\begin{eqnarray}\label{nbin}
&&\pi(s_{i}|\lambda_{i},\tau)\nonumber
\\[-8pt]\\[-8pt]
&&\quad=\frac{\Gamma(s_{i}+\tau)}{\Gamma(s_{i}+1)\Gamma(\tau)}
\biggl(\frac{\tau}{\tau+\lambda_{i}}\biggr)^{\tau}\biggl(\frac{\lambda_{i}}{\tau+\lambda_{i}}
\biggr)^{s_{i}},\nonumber
\end{eqnarray}
for $s_{i} \in \mathbb{N}\cup \{0\}$, with the same mean as the Poisson
regression model, that is, $\mathbb{E}(s_{i}) = \lambda_{i}$, and
$\operatorname{Var}(s_{i})=\lambda_{i}+\lambda_{i}^{2}/\tau$,
with the added parameter $\tau$ capturing the variance inflation
associated with over-dispersion.

\subsection{Survival Data} \label{gpcox}

The modeling approach via Gaussian processes~ex\-ploited above extends to
other classes of models, for example, those for survival data. In
survival studies the task is typically to measure the effect~of a~set~%
of variables on the survival time, that is, the time to a particular
event or ``failure'' of interest, such as death or occurrence of a
disease. The Cox proportio\-nal hazard model of \citet{cox1972} is an
extremely popular choice. The model is defined through the hazard rate
function $h(t|\mathbf{x}_i)=h_0(t)\exp(\mathbf{x}_i'\bolds{\beta})$,
where $h_0(\cdot)$ is the baseline hazard function, $t$ is the failure
time and $\bolds{\beta}$ the $p$-dimensional regression coefficient
vector. The cumulative baseline hazard function is denoted as
$H_0(t)=\int_0^th_0(u)\,du$ and the survivor function becomes
$S(t|\mathbf{x}_i)=S_{0}(t)^{\exp(\mathbf{x}_i'\bolds{\beta})}$, where
$S_{0}(t) = \exp\{-H_0(t)\}$ is the baseline survivor function.

Let us indicate the data as
$(t_1,\mathbf{x}_1,d_1),\ldots,(t_n,\mathbf{x}_n,\break d_n)$ with censoring
index $d_i=0$ if the observation is right censored and $d_i=1$ if the
failure time $t_i$ is observed. A GP model for survival data is defined~as
\begin{equation}\label{modelCox}
\hspace*{12pt}h(t_i|z(\mathbf{x}_i))=h_0(t_i)\exp(z(\mathbf{x}_i)),\quad
i=1,2,\ldots,n,\hspace*{-12pt}
\end{equation}
with $\mathbf{z}(\mathbf{X})$ as in (\ref{ZvecDist}). In this general setting,
defining a probability model for Bayesian analysis requires the
identification of a prior formulation for the cumulative baseline
hazard function. One strategy often adopted in the literature on
survival models is to utilize the partial likelihood of \citet{cox1972}
that avoids prior specification and estimation of the baseline hazard,
achieving a parsimonious representation of the model. Alternatively,
\citet{kalb1978} employs a nonparametric gamma process prior on
$H_0(t_i)$ and then calculates a marginalized likelihood. This ``full''
likelihood formulation tends to behave similarly to the partial
likelihood one when the concentration parameter of the gamma process
prior tends to $0$, placing no confidence in the initial parametric
guess. \citet{sinha2003} extend this theoretical justification to
time-dependent covariates and time-varying regression parameters, as
well as to grouped survival data.

\section{Choice of the GP Covariance Matrix}\label{GPcov}

We explore alternative covariance formulations for the Gaussian process
prior (\ref{ZvecDist}) and demonstrate the flexibility of the
construction. In general, any plausible relationship between the
covariates and the response can be represented through the choice of
$\mathbf{C}$, as long as the condition of positive definiteness of the
matrix is satisfied; see \citet{thrun2004}. In the
\hyperref[app]{Appendix} we further address practical computational
issues that arise in the application of Gaussian processes due to
numerical instability in the construction of the covariance matrix and
the calculation of its inverse.

\subsection{$1$-term vs. $2$-term Exponential Forms}

We consider covariance functions that include a constant term and a
nonlinear, exponential term as
\begin{equation}\label{covmatrix}
\mathbf{C}=\operatorname{Cov}(\mathbf{z}(\mathbf{X}))=\frac{1}{\lambda_{a}}\mathbf{J}_n
+\frac{1}{\lambda_{z}}\exp(-\mathbf{G}),
\end{equation}
with $\mathbf{J}_n$ an $n \times n$ matrix of $1$'s and
$\exp(\mathbf{G})$ a matrix with elements $\exp(g_{ij})$, where
$g_{ij}=(\mathbf{x}_{i}-\mathbf{x}_{j})'
\mathbf{P}(\mathbf{x}_{i}-\mathbf{x}_{j})$ and
$\mathbf{P}=\operatorname{diag}(-\log(\rho_{1},\ldots,\rho_{p}))$, with
$\rho_{k} \in [0,1]$ associated to $x_{k}, k=1,\ldots,p$. In the
literature on Gaussian processes a noise component, called ``jitter,''
is sometimes added to the covariance matrix $\mathbf{C}$, in addition
to the term $(1/\lambda)\mathbf{J}$, in order to make the matrix
computations better conditioned; see \citet{neal1999}. This is
consistent with the belief that there may be unexplained sources of
variation in the data, perhaps due to explanatory variables that were
not recorded in the original study. The parametrization of $\mathbf{G}$
we adopt allows simpler prior specifications (see below), and it is
also used by \citet{link2006} as a transformation of the exponential
term used by \citet{neal1999} and \citet{sacks2000} in their
formulations.  Neal (1999) notices that introducing an intercept in
model (\ref{uni}), with precision parameter $\lambda_a$, placing a
Gaussian prior on it and then marginalizing over the intercept produces
the additive covariance structure (\ref{covmatrix}). The parameter for
the exponential term, $\lambda_z$, serves as a scaling factor for this
term. In our empirical investigations we found that construction
(\ref{covmatrix}) is sensitive to scaling and that best results can be
obtained by normalizing $\mathbf{X}$ to lie in the unit cube,
$[0,1]^{p}$, though standardizing the columns to mean $0$ and variance
$1$ produces similar results.

\begin{figure*}

\includegraphics{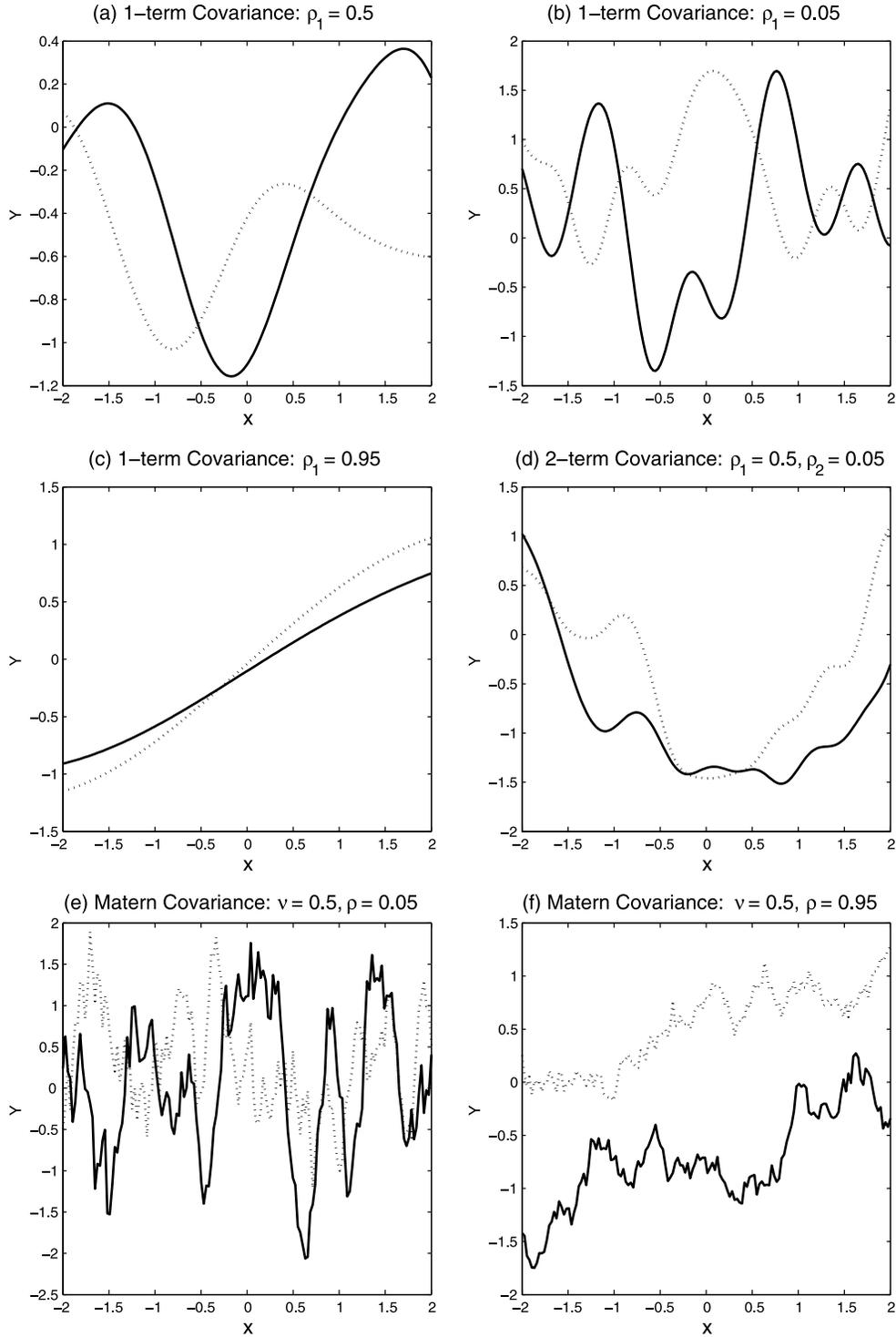}

\caption{Response curves drawn from a GP. Each plot shows two (solid
and dashed) random realizations. Plots \textup{(a)--(c)} were obtained with the
exponential covariance (\protect\ref{covmatrix}) and plot \textup{(d)} with the 2-term
formulation (\protect\ref{covmatrix2}). Plots \textup{(e)} and \textup{(f)} show realizations from
the matern construction. All curves employ a one-dimensional
covariate.} \label{GPcurves}
\end{figure*}

The single-term exponential covariance provides a parsimonious
representation that enables a broad class of linear and nonlinear
response surfaces. Plots (a)--(c) of Figure \ref{GPcurves} show response
curves produced by utilizing a GP with the exponential covariance
matrix~(\ref{covmatrix}) and three different values of $\rho$. One
readily notes how higher order polynomial-type response surfaces can be
generated by choosing relatively lower values for $\rho$, whereas the
assignment of higher values provides lower order polynomial-type that
can also include roughly linear response surfaces [plot (c)].

We also consider a two-term covariance obtained by adding a second
exponential term to (\ref{covmatrix}), that is,
\begin{eqnarray}\label{covmatrix2}
\hspace*{16pt}\mathbf{C}&=&\operatorname{Cov}(\mathbf{z}(\mathbf{X}))\nonumber\hspace*{-16pt}
\\[-5pt]\\[-5pt]
&=&\frac{1}{\lambda_{a}}\mathbf{J}_n + \frac{1}{\lambda_{1,z}} \exp(-\mathbf{G}_{1})
+\frac{1}{\lambda_{2,z}}\exp(-\mathbf{G}_{2}),\nonumber
\vspace*{4pt}
\end{eqnarray}
where $\mathbf{G}_{1}$ and $\mathbf{G}_{2}$ are parameterized as
$\mathbf{P}_{1}\break
=\operatorname{diag}(-\log(\rho_{1,1},\ldots,\rho_{1,p}))$ and
$\mathbf{P}_{2} =\operatorname{diag}
(-\log(\rho_{2,1},\break\ldots,\rho_{2,p}))$, respectively. As noted in
\citet{neal2000}, ad\-ding multiple terms results in rougher, more
complex, surfaces while retaining the relative computational efficiency
of the exponential formulation. For example, plot (d) of Figure
\ref{GPcurves} shows examples of surfaces that can be generated by
employing the $2$-term covariance formulation with $(\rho_{1} ,
\rho_{2}) = (0.5, 0.05)$ and $(\lambda_{1,z} = 1, \lambda_{2,z} =
8)$.

\subsection{The Matern Construction}

An alternative choice to the exponential covariance term is the Matern
formulation. This introduces an explicit smoothing parameter, $\nu$,
such that the resulting Gaussian process is $k$ times differentiable
for $k \leq \nu$,
\begin{eqnarray}\label{matern}
&&\hspace*{7pt}\mathbf{C}(z(\mathbf{x}_{i}), z(\mathbf{x}_{j}))\hspace*{-7pt} \nonumber
\\[-8pt]\\[-8pt]
&&\hspace*{7pt}\quad= \frac{1}{2^{\nu-1}\Gamma(\nu)}\bigl[2\sqrt{\nu d(\mathbf{x}_{i},\mathbf{x}_{j})}\bigr]^{\nu} K_{\nu}\bigl[2\sqrt{\nu d(\mathbf{x}_{i},\mathbf{x}_{j})}\bigr],\hspace*{-7pt}\nonumber
\end{eqnarray}
with $d(\mathbf{x}_{i},\mathbf{x}_{j}) =
(\mathbf{x}_{i}-\mathbf{x}_{j})'\mathbf{P}(\mathbf{x}_{i}-\mathbf{x}_{j})$,
$K_{\nu}(\cdot)$ the Bes\-sel function and $\mathbf{P}$ parameterized as
in (\ref{covmatrix}). \citet{banerjee2008} employ such a construction
with $\nu$ fixed to $0.5$ for modeling a spacial random effects process
characterized by roughness.  One recovers the exponential covariance
term from the Matern construction in the limit as $\nu \rightarrow
\infty$. However, \citet{rasm2006} point out that two formulations are
essentially the same for $\nu > \frac{7}{2}$, as confirmed by our own
simulations.

\section{Prior Model for Bayesian Variable Selection}\label{priors}

The unified modeling approach we have described allows
us to put forward a general framework for variable selection that
employs Bayesian methods and mixture priors for the selection of the
predictors. In particular, variable selection can be achieved within
the GP modeling framework by imposing ``spi\-ke-and-slab'' mixture priors
on the covariance parameters in (\ref{covmatrix}), that is,
\begin{equation}\label{jointvsprior}
\hspace*{12pt}\pi(\rho_{k}|\gamma_{k}) = \gamma_{k}\mathbb{I}[0 \leq \rho_{k} \leq 1]
+ (1-\gamma_{k})\delta_{1}(\rho_{k}),\hspace*{-12pt}
\end{equation}
for $k = 1, \ldots, p$, with $\delta_1(\cdot)$ a point mass
distribution at one. Clearly, $\rho_k=1$ causes the predictor $x_k$ to
have no effect on the computation for the GP covariance matrix. This
formulation is similar in spirit to the use of selection priors for
linear regression models and is employed by \citet{link2006} in the
univariate GP regression framework (\ref{uni}). Further Bernoulli
priors are imposed on the selection parameters, that is, $\gamma_k \sim
\operatorname{Bernoulli}(\alpha_k)$ and Gamma priors are specified on the
precision terms $(\lambda_{a},\lambda_{z})$.

Variable selection with a covariance matrix that employs two
exponential terms as in (\ref{covmatrix2}) is more complex.  In
particular, one can select covariates separately for each exponential
term by assigning a~specific set of variable selection parameters to
each term, that is, $(\bolds{\gamma}_{1},\bolds{\gamma}_{2})$ associated to
$(\bolds{\rho}_{1},\bolds{\rho}_{2})$, and simply extending the single term
formulation via independent spike-and-slab priors of the form
\begin{eqnarray}
&&\label{sepvsprior1}\hspace*{10pt}\pi(\rho_{1,k}|\gamma_{1,k})\hspace*{-10pt} \nonumber
\\[-8pt]\\[-8pt]
&&\hspace*{10pt}\quad= \gamma_{1,k}\mathbb{I}[0
\leq\rho_{1,k} \leq 1] +(1-\gamma_{1,k}) \delta_{1}(\rho_{1,k}),\hspace*{-10pt}\nonumber
\\
&&\label{sepvsprior2}\hspace*{10pt}\pi(\rho_{2,k}|\gamma_{2,k})\hspace*{-10pt}\nonumber
\\[-8pt]\\[-8pt]
 &&\hspace*{10pt}\quad= \gamma_{2,k}\mathbb{I}[0
\leq\rho_{2,k} \leq 1] +(1-\gamma_{2,k}) \delta_{1}(\rho_{2,k}),\hspace*{-10pt}\nonumber
\end{eqnarray}
with $k = 1,\ldots,p$.  Assuming \emph{a priori} independence of the
two model spaces, Bernoulli priors can be imposed on the selection
parameters, that is, $\gamma_{i,k} \sim \operatorname{Bernoulli}(\alpha_{i,k}),
i= 1,2$. This variable selection framework identifies the association
of each covariate, $x_{k}$, to one or both terms. Final selection can
then be accomplished by choosing the covariates in the union of those
selected by \textit{either} of the two terms. An alternative strategy for
variable selection may employ a common set of variable selection
parameters, $\bolds{\gamma} = (\gamma_{1},\ldots,\gamma_{p})$ for
\textit{both} $\bolds{\rho}_{1}$ and $\bolds{\rho}_{2}$, in a joint
spike-and-slab (product) prior formulation,
\begin{eqnarray}\label{jointvsprior2}
&&\pi(\rho_{1,k},\rho_{2,k}|\gamma_{k})\nonumber \\
&&\quad=
\gamma_{k}\mathbb{I}[0 \leq\rho_{1,k} \leq 1]\mathbb{I} [0
\leq\rho_{2,k} \leq 1]
\\
&&\qquad{}
+(1-\gamma_{k})\delta_{1}(\rho_{1,k})\delta_{1}(\rho_{2,k}),\nonumber
\end{eqnarray}
where we assume a priori independence of the parameter spaces,
$\bolds{\rho}_{1}$ and $\bolds{\rho}_{2}$. This prior choice focuses more on
overall covariate selection, rather than simultaneous selection and
assignment to each term in (\ref{covmatrix2}).  While we lose the
ability to align the $\rho_{i,k}$ to each covariance function term, we
expect to improve computational efficiency by jointly sampling
$(\bolds{\gamma},\bolds{\rho}_{1},\bolds{\rho}_{2})$ at each iteration of the
MCMC scheme as compared to a separate joint sampling on
$(\bolds{\gamma}_{1},\bolds{\rho}_{1})$ and
$(\bolds{\gamma}_{2},\bolds{\rho}_{2})$. Some investigation is done in
\citet{terrance2010}.

\section{Posterior Inference}\label{posteriors}

The methods for posterior inference we are going to describe apply to
all GP formulations, even though we focus our simulation work on the
continuous and count data models.  We therefore express the posterior
formulation employing a generalized notation. First, we collect all
parameters of the GP covariance matrix in $\bolds\Theta$ and write
$\mathbf{C}=\mathbf{C}(\bolds{\Theta})$. For example, for covariance
matrix of type (\ref{covmatrix}) we have
$\bolds{\Theta}=(\bolds{\rho},\lambda_{a},\lambda_{z})$. Next, we
extend our notation to include the selection parameter $\bolds{\gamma}$
by using
$\bolds{\Theta}_{\bolds{\gamma}}=(\bolds{\rho}_{\bolds{\gamma}},\lambda_{a},\lambda_{z})$
to indicate that $\rho_{k}=1$ when $\gamma_{k} = 0$, for
$k=1,\ldots,p$. For covariance of type (\ref{covmatrix2}) we write
$\bolds{\Theta}_{\bolds{\gamma}}=\{\bolds{\Theta}_{\bolds{\gamma}_{1}},
\bolds{\Theta}_{\bolds{\gamma}_{2}}, \lambda_{a}\}$, where
$\bolds{\gamma} = (\bolds{\gamma}_{1},\bolds{\gamma}_{2})'$ and
$\bolds{\Theta}_{\bolds{\gamma}_{i}}=(\bolds{\rho}_{i\bolds{\gamma}_{i}},
\lambda_{i,z}),i \in\{1,2\}$ for prior of type
(\ref{sepvsprior1})--(\ref{sepvsprior2}) and
$\bolds{\Theta}_{\bolds{\gamma}}=(\bolds{\rho}_{1\bolds{\gamma}},\bolds{\rho}_{2\bolds{\gamma}},\lambda_{a},\lambda_{1,z},\lambda_{2,z})$
for prior of type (\ref{jointvsprior2}),  and similarly for the Matern
construction. Next, we define
$D_{i}\in\{y_{i},\{s_{i},z(\mathbf{x}_{i})\}\}$ and $\mathbf{D} :=
\{D_{1},\ldots,D_{n}\}$ to capture the observed data \textit{augmen\-ted}
by the unobserved GP variate, $\mathbf{z}(\mathbf{X})$, for the latent
response models [such as model (\ref{nbin}) for count data]. Finally,
we set $\mathbf{h}:=\{r,\tau\}$ to group unique parameters $\notin
\bolds{\Theta}_{\bolds{\gamma}}$ and we collect hyperparameters in
$\mathbf{m}:=\{\mathbf{a},\mathbf{b}\}$, with $\mathbf{a} =
\{a_{\lambda_a}, a_{\lambda_z}, a_{r}, a_{\tau}\}$ and similarly for
$\mathbf{b}$, where $\mathbf{a}$ and $\mathbf{b}$ include the shape and
rate hyperparameters of the Gamma priors on the associated parameters.
With this notation we can finally outline a generalized expression for
the full conditional of
$(\bolds{\gamma},\bolds{\rho}_{\bolds{\gamma}})$ as
\begin{eqnarray}\label{conditional}
&&\pi(\bolds{\gamma},\bolds{\rho}_{\bolds{\gamma}}|\bolds{\Theta}_{\bolds{\gamma}}\backslash
\bolds{\rho}_{\bolds{\gamma}},\mathbf{D},\mathbf{h},\mathbf{m})\nonumber
\\[-8pt]\\[-8pt]
&&\quad\propto L^{a}(\bolds{\gamma},
\bolds{\rho}_{\bolds{\gamma}}|\bolds{\Theta}_{\bolds{\gamma}}\backslash\bolds{\rho}_{\bolds{\gamma}},\mathbf{D},\mathbf{h},\mathbf{m})\pi(\bolds{\gamma}),\nonumber
\end{eqnarray}
with $L^{a}$ the augmented likelihood. Notice that the term
$\pi(\bolds{\rho}_{\bolds{\gamma}}|\bolds{\gamma})$ does not appear in
(\ref{conditional}) since $\pi(\rho_{k}|\break \gamma_{k}) = 1$, for
$k=1,\ldots, p$.

\subsection{Markov Chain Monte Carlo---Scheme 1}\label{MHMCMC}

We first describe a Metropolis--Hastings scheme within Gibbs sampling
to jointly sample $(\bolds{\gamma},\bolds{\rho}_{\bolds{\gamma}})$,
which is an adaptation of the MCMC model comparison ($\mathrm{MC}^{3}$)
algorithm originally outlined in \citet{madigan1995} and extensively
used in the variable selection literature. As we are unable to
marginalize over the parameter space, we need to modify the algorithm
in a hierarchical fashion, using the move types outlined below.
Additionally, we need to sample all the other nuisance parameters.

A generic iteration of this MCMC procedure comprises the following
steps:
\begin{longlist}[(2)]
 \item[(1)] \textit{Update} $({\gamma},{\rho}_{{\gamma}})$: Randomly choose
among three between-models transition moves:
    \begin{longlist}[{(iii)}]
    \item[{(i)}] Add: set $\gamma_{k}' = 1$ and sample $\rho_k'$ from a $\mathcal{U}(0,1)$ proposal. Position $k$ is randomly chosen from the set of $k$'s where $\gamma_{k}=0$ at the previous iteration.
    \item[{(ii)}] Delete: set $(\gamma_{k}' = 0, \rho_{k}'=1)$. This results in covariate $x_{k}$ being excluded in the current iteration. Position $k$ is randomly chosen from among those included in the model at the previous iteration.
    \item[{(iii)}] Swap: perform both an \emph{Add} and \emph{Delete} move.  This move type helps to more quickly traverse a~large covariate space.
    \end{longlist}
   The proposed value $(\bolds{\gamma}',\bolds{\rho}_{\bolds{\gamma}'}')$
is accepted with probability,
\[
\alpha =
\min\biggl\{1,\frac{\pi(\bolds{\gamma}',\bolds{\rho}_{\bolds{\gamma}'}'|\bolds{\Theta}_{\bolds{\gamma}'}\backslash\bolds{\rho}'_{\bolds{\gamma}'},
\mathbf{D},\mathbf{h},\mathbf{m})q(\bolds{\gamma}|\bolds{\gamma}')}{\pi(\bolds{\gamma},\bolds{\rho}_{\bolds{\gamma}}|\bolds{\Theta}_{\bolds{\gamma}}\backslash\bolds{\rho}_{\bolds{\gamma}},
\mathbf{D},\mathbf{h},\mathbf{m})q(\bolds{\gamma}'|\bolds{\gamma})}\biggr\},
\]
where the ratio of the proposals
$q(\bolds{\rho}_{\bolds{\gamma}})/q(\bolds{\rho}_{\bolds{\gamma}'}')$
drops out of the computation since we employ a $\mathcal{U}(0,1)$
proposal.
\item[(2)] Execute a Gibbs-type move, \emph{Keep}, by sampling from a
$\mathcal{U}(0,1)$ all $\rho'_k$'s such that $\gamma_{k}'=1$.
This move is not required for ergodicity, but it allows to perform a
refinement of the parameter space within the existing model, for faster
convergence.
\item[(3)] \textit{Update} $\{\lambda_{a},\lambda_{z}\}$: These are
updated using Me\-tropolis--Hastings moves with Gamma proposals centered
on the previously sampled values.
\item[(4)] \textit{Update} $\mathbf{h}$: Individual model parameters in
$\mathbf{h}$ are updated using Metropolis--Hastings moves with proposals
centered on the previously sampled values.
\item[(5)] \textit{Update} $\mathbf{z}$: Jointly sample $\mathbf{z}$
for latent response models using the approach enumerated in
\citet{neal1999} with proposal $\mathbf{z}' = (1
-\varepsilon^{2})^{1/2}\mathbf{z} +
\varepsilon\mathbf{L}\mathbf{u}$, where $\mathbf{u}$ is a~vector of
i.i.d. standard Gaussian values and $\mathbf{L}$ is the Cholesky
decomposition of the GP covarian\-ce matrix. For faster convergence $R$
consecutive updates are performed at each iteration.
\end{longlist}

\citet{green1995} introduced a Markov chain Monte Carlo method for
Bayesian model determination for the situation where the dimensionality
of the parameter vector varies iteration by iteration. Recently,
\citet{gottardo2008} have shown that the reversible jump can be
formulated in terms of a mixture of singular distributions. Following
the results given in their examples, it is possible to show that the
acceptance probability of the reversible jump~for\-mulation is the same
as in the Metropolis--Hastings algorithm described above, and therefore
that the two algorithms are equivalent; see \citet{terrance2010}.

For inference, estimates of the marginal posterior probabilities of
$\gamma_k=1$, for $k=1\ldots,p$, can be computed based on the MCMC
output. A simple strategy is to compute Monte Carlo estimates by
counting the number of appearances of each covariate across the visited
models. Alternatively, Rao--Blackwellized estimates can be calculated by
averaging the full conditional probabilities of $\gamma_k=1$. Although
computationally more expensive, the latter strategy may result in
estimates with better precision, as noted by \citet{guan2009}. In all
simulations and examples reported below we obtained satisfactory
results by estimating the marginal posterior probabilities by counts
restricted to be\-tween-models moves, to avoid over-estimation.

\subsection{Markov Chain Monte Carlo---Scheme 2} \label{2StepGibbs}

Next we enumerate a Markov chain Monte Carlo algorithm to directly
sample $(\bolds{\gamma},\bolds{\rho}_{\bolds{\gamma}})$ with a Gibbs scan that
employs a Metropolis acceptance step.  We formulate a proposal
distribution of a similar mixture form as the joint posterior by
extending a result from \citet{gottardo2008} to produce a~move to
$(\gamma_{k}=0,\rho_{k}=1)$, as well as to $(\gamma_{k}=1,\rho_{k}
= [0,1))$.

A generic iteration of this MCMC procedure comprises the following steps:
\begin{longlist}[(2)]
\item[(1)] For $k = 1,\ldots,p$ perform a joint update for
($\gamma_{k},\rho_{k})$ with two moves, conducted in succession:
\begin{longlist}[{(ii)}]
\item[{(i)}] Between-models: Jointly propose a new model such
that if $\gamma_{k}=1$, propose $\gamma_{k}'=0$ and set
$\rho_{k}'=1$; otherwise, propose  $\gamma_{k}'=1$ and draw
$\rho_{k}' \sim \mathcal{U}(0,1)$. Accept the proposal for
$(\gamma_{k}',\rho_{k}')$ with probability,
\[
\alpha =
\min\biggl\{1,\frac{\pi(\gamma_{k}',\rho_{k}'|\bolds{\gamma}_{(k)}',
\bolds{\Theta}_{\bolds{\gamma}_{(k)}'},\mathbf{D},\mathbf{h},\mathbf{m})}
{\pi(\gamma_{k},\rho_{k}|\bolds{\gamma}_{(k)}',\bolds{\Theta}_{\bolds{\gamma}_{(k)}'},
\mathbf{D},\mathbf{h},\mathbf{m})}\biggr\},
\]
where now $\bolds{\gamma}_{(k)}' : =
(\gamma_{1}',\ldots,\gamma_{k-1}',
\gamma_{k+1},\ldots,\gamma_{p})$ and si\-milarly for
$\bolds{\rho}_{(k)}' \in
\bolds{\Theta}_{\bolds{\gamma}_{(k)}'}$. The joint proposal ratio
for $(\gamma_{k},\rho_{k})$, reduces to $1$ since we employ a
$\mathcal{U}(0,1)$ proposal for $\rho_{k} \in [0,1]$ and a symmetric
Dirac measure proposal for $\gamma_{k}$.
\item[{(ii)}] Within
model: This move is performed only if we sample $\gamma_{k}'=1$ from
the between-models move, in which case we propose $\gamma_{k}''=1$
and, as before, draw $\rho_{k}''\sim \mathcal{U}(0,1)$. Similar to
the between-models move, accept the joint proposal for
$(\gamma_{k}'',\rho_{k}'')$ with probability,
\[
\alpha =
\min\biggl\{1,\frac{\pi(\gamma_{k}'',\rho_{k}''|\bolds{\gamma}_{(k)}',\bolds{\Theta}_{\bolds{\gamma}_{(k)}'},\mathbf{D},\mathbf{h},
\mathbf{m})}{\pi(\gamma_{k}',\rho_{k}'|\bolds{\gamma}_{(k)}',
\bolds{\Theta}_{\bolds{\gamma}_{(k)}'},\mathbf{D},\mathbf{h},\mathbf{m})}\biggr\},
\]
which further reduces to just the ratio of posteriors since we propose
a move within the current model and utilize a $\mathcal{U}(0,1)$
proposal for $\rho_{k}$.
\end{longlist}
\item[(2)] Sample the parameters
$\{\lambda_{a},\lambda_{z},\mathbf{h}\}$ and latent responses
$\mathbf{z}$ as outlined in scheme 1.
\end{longlist}

In simulations we also investigate performances of an adaptive scheme
that employs a proposal with tuning parameters adapted based on
``learning'' from the data. In particular, we employ the method of
\citet{haario2001} for our Ber\-noulli proposal for $\gamma|\alpha$ to
successively update the mean parameter, $\alpha_{k},k=1,\ldots,p$,
based on prior sampled values for $\gamma_{k}$. The construction does
not require additional likelihood computations and it is expected to
achieve more rapid convergence in the model space than the nonadaptive
scheme. \citet{roberts2007} and \citet{ji2009} note conditions under
which adaptive schemes achie\-ve convergence to the target posterior
distribution.

Schemes 1 and 2 we enumerated above may be easily modified when
employing the $2$-term covariance formulation (\ref{covmatrix2}); see
\citet{terrance2010}.

\subsection{Prediction}

Let ${\mathbf{z}_{f}}=\mathbf{z}(\mathbf{X}_{f})$ be an $n_{f} \times
1$ latent vector of future cases.  We use the regression model
(\ref{uni}) to demonstrate prediction under the GP framework. The joint
distribution over training and test sets is defined to be
$\mathbf{z}_{*}:=[\mathbf{z}', \mathbf{z}_f']' \sim
\mathcal{N}(\mathbf{0},\mathbf{C}_{n +n_f})$ with covariance,
\[
\mathbf{C}_{n+n_{f}} :=
\pmatrix{
\mathbf{C}_{(\mathbf{X},\mathbf{X})} &
\mathbf{C}_{(\mathbf{X},\mathbf{X}_{f})} \cr
\mathbf{C}_{(\mathbf{X}_{f},\mathbf{X})} &
\mathbf{C}_{(\mathbf{X}_{f},\mathbf{X}_{f})}},
\]
where
$\mathbf{C}_{(\mathbf{X},\mathbf{X})}:=\mathbf{C}_{(\mathbf{X},\mathbf{X})}(\bolds{\Theta})$.
The conditional joint predictive distribution over the test cases,
$\mathbf{z}_{f}|\mathbf{z}$, is also multivariate normal distribution
with expectation
$\mathbb{E}[\mathbf{z}_{f}|\mathbf{z}]=\mathbf{C}_{(\mathbf{X}_{f},\mathbf{X})}
\mathbf{C}_{(\mathbf{X},\mathbf{X})}^{-1} \mathbf{z}$. Estimation is
based on the posterior MCMC samples. Here we take a~computationally
simple approach by first estimating $\hat{\mathbf{z}}$ as the mean of
all sampled values of $\mathbf{z}$, defining
\begin{equation}\label{predict}
\mathbf{D}(\bolds{\Theta}) :=
\mathbf{C}_{(\mathbf{X}_f,\mathbf{X})}\mathbf{C}_{(\mathbf{X},\mathbf{X})}^{-1}\hat{\mathbf{z}},
\end{equation}
and then estimating the response value as
\begin{equation}\label{predict2}
\hat{\mathbf{y}}_{f} = \hat{\mathbf{z}}_{f}|\hat{\mathbf{z}} =
\frac{1}{K} \sum_{t=1}^{K}\mathbf{D}(\bolds{\Theta}^{(t)}),
\end{equation}
with $K$ the number of MCMC iterations and where calculations of the
covariance matrices in (\ref{predict}) are restricted to the variables
selected based on the margi\-nal posterior probabilities of $\gamma_k=1$.
A more coherent estimation procedure, that may return more precise
estimates but that is also computationally more expensive, would
compute Rao--Blackwellized estimates by averaging the predictive
probabilities over all visited models; see \citet{guan2009}. In the
simulations and examples reported below we have calculated
(\ref{predict2}) using every 10th MCMC sampled value, to provide a
relatively less correlated sample and save on computational time. In
addition, when computing the variance product term in (\ref{predict}),
we have employed the Cholesky decomposition $\mathbf{C} =
\mathbf{L}\mathbf{L}'$, following \citet{neal1999}, to avoid direct
computation of the inverse of $\mathbf{C}_{(\mathbf{X},\mathbf{X})}$.

For categorical data models, we may predict the new class labels,
$\mathbf{t}_{f}$, via the rule of largest probability in the case of
a~binary logit model, with estimated latent realizations
$\hat{\mathbf{z}}_{f}$, and via data augmentation based on the values
of $\hat{\mathbf{y}}_{f}$ in the case of a~binary probit model.

\subsubsection{Survival Function Estimation}

For survival data it is of interest to estimate the survivor function
for a new subject with unknown event time, $T_{i}$, and associated
$z_{f,i} := z_{f,i}(\mathbf{x}_{f,i})$. This is defined as
\begin{eqnarray}\label{survivor}
P(T_{i}\geq
t|z_{f,i},\mathbf{z})&=&S_{i}(t|z_{f,i},\mathbf{z})\nonumber
\\[-8pt]\\[-8pt]
&=& S_{0}(t|\mathbf{z})^{\exp(z_{f,i})}.\nonumber
\end{eqnarray}
When using the partial likelihood formulation an empirical Bayes
estimate of the baseline survivor function, $S_{0}(t|\mathbf{z})$, must
be calculated, since the model does not specifically enumerate the
baseline hazard. \citet{weng2007}, for example, propose a method that
discretizes the likelihood to produce an estimator with the useful
property that it cannot take negative values. Accuracy of this estimate
may be potentially improved by Rao--Blackwellizing the computation by
averaging over the MCMC runs.

\begin{figure*}

\includegraphics{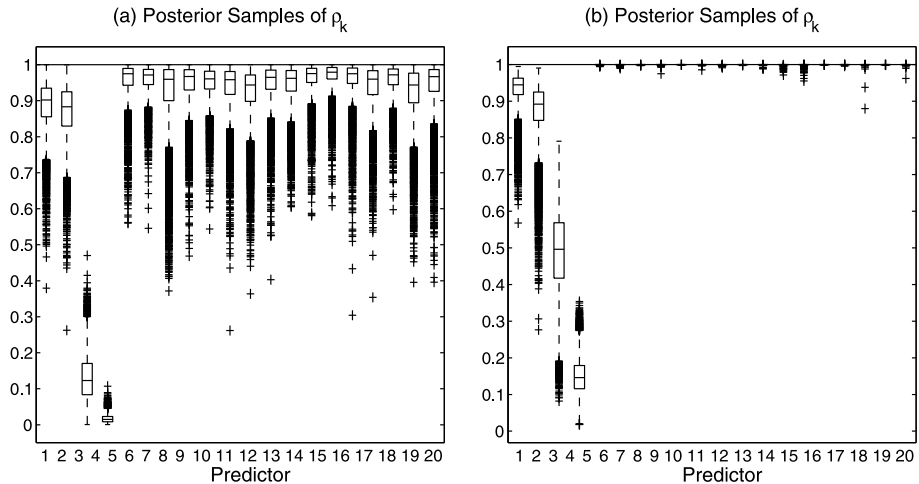}

\caption{Use of variable selection parameters: Simulated data $(n=80, p
= 20)$. Box plots of posterior samples for $\rho_{k} \in [0,1]$. Plots
(\textup{a}) and (\textup{b}) demonstrate selection without and with, respectively, the
inclusion of the selection parameter $\bolds{\gamma}$.}\label{link}
\end{figure*}

\section{Simulation Study}\label{Simulation}

\subsection{Parameter Settings}\label{sse:hyper}

In all simulations and applications reported in this paper we set both
priors on $\lambda_a$ and $\lambda_z$ as $\mathcal{G}(1,1)$. We did not
observe any strong sensitivity to this choice. In particular, we
considered different choices of the two parameters of these Gammma
priors in the range $(0.01,1)$, keeping the prior mean at 1 but with
progressively larger variances, and observed ve\-ry little change in the
range of posterior sampled values. We also experimented with prior mean
values of $10$ and $100$, which produced only a small impact on the
posterior. For model (\ref{uni}) we set $r \sim
\mathcal{G}(a_{r},b_{r})$ with $(a_{r}, b_{r}) = (2,0.1)$ to reflect
our a~priori expected residual variance. For the count model
(\ref{nbin}), we set $\tau\sim\mathcal{G}(1,1)$. For survival data, when
using the full likelihood from \citet{kalb1978} we specified a $\mathcal{G}(1,1)$
prior for both the parameter of the exponential base
distribution and the concentration parameter of the Gamma process prior
on the baseline.

Some sensitivity on the Bernoulli priors on the $\gamma_k$'s is, of
course, to be expected, since these priors drive the sparsity of the
model. Generally speaking, parsimonious models can be selected by
specifying $\gamma_k\sim  \operatorname{Bernoulli}(\alpha_{k})$ with
$\alpha_k=\alpha$ and $\alpha$ a small percentage of the total number
of variables. In our simulations we set $\alpha_{k}$ to $0.025$. We
observed little sensitivity in the results  for small changes around
this value, in the range of 0.01--0.05, though we would expect to see
significant sensitivity for much higher values of $\alpha$. We also
investigated sensitivity to a~Beta hyperprior on $\alpha$; see below.

When running the MCMC algorithms independent chain samplers with
$\mathcal{U}(0,1)$ proposals for the $\rho_{k}$'s have worked well in
all applications reported in this paper, where we have always
approximately achieved the target acceptance rate of 40--60\%
indicating efficient posterior sampling.

\subsection{Use of Variable Selection Parameters}

We first demonstrate the advantage of introducing selection parameters
in the model. Figure \ref{link} shows results with and without the
inclusion of the variable selection parameter vector $\bolds{\gamma}$ on a
simulated scenario with a kernel that incorporates both linear and
nonlinear associations. The observed continuous response, $y$, is
constructed from a mix of linear and nonliner relationships to $4$
variables, each generated from a $\mathcal{U}(0,1)$,
\[
y = x_{1} + x_{2} + \sin(3x_{3}) + \sin(5x_{4})+
\varepsilon,
\]
with $\varepsilon \sim \mathcal{N}(0,\sigma^{2})$ and $\sigma = 0.05$.
Additional variables are randomly generated, again from
$\mathcal{U}(0,1)$. In this simulation we used $(n,p) = (80,20)$. We
ran 70,000 MCMC iterations, of which 10,000 were discarded as burn-in.

Plot (a) of Figure \ref{link} displays box plots of the MCMC samples
for the $\rho_{k}'s, k = 1, \ldots ,20$, for the case of no variable
selection, that is, by using a simple ``slab'' prior on the $\rho_k$'s.
As both \citet{link2006} and \citet{neal2000} note, the single
covariates demonstrate an association to the response whose strength
may be assessed utilizing the distance of the posterior samples of the
$\rho_k$'s from $1$. One notes that, according to this criterion, the
true covariates are all selected. It is conceivable, however, for some
of the unrelated covariates to be selected using the same criterion,
since the $\rho_{k}$'s all sample below $1$, and that this problem
would be compounded as $p$ grows. Plot (b) of Figure \ref{link},
instead, captures results from employing the variable selection
parameters $\bolds{\gamma}$ and shows how the inclusion of these
parameters results in the sampled values of the $\rho_{k}$'s for
variables unrelated to the response being all pushed up against
$1$.\looseness=-1

This simple simulated scenario also helps us to illustrate a couple of
other features. First, a single exponential term in (\ref{covmatrix})
is able to capture a wide variety of continuous response surfaces,
allowing a great flexibility in the shape of the response surface, with
the linear fit being a subset of one of many types of surfaces that can
be generated. Second, the effect of covariates with higher-order
polynomial-like association to the response is captured by having
estimates of the corresponding $\rho_k$'s further away from~1; see, for
example, covariate $x_{4}$ in Figure \ref{link} which expresses the
highest order association to the response.

\begin{table*}
\caption{Large $p$: Simulations for continuous, count and survival data
models with $(n,p) = (100,1{,}000)$}\label{table:simresults}
\begin{tabular*}{345pt}{@{\extracolsep{\fill}}lccc@{}}
\hline
& \textbf{Continuous data} & \textbf{Count data} & \textbf{Cox model}  \\
\hline
Coefficients:& & & \\
$a_{1}$ & $1.0$ & $1.6$ & \phantom{$-$}$3.0$\\
$a_{2}$ & $1.0$ & $1.6$ & $-2.5$\\
$a_{3}$ & $1.0$ & $1.6$ & \phantom{$-$}$3.5$\\
$a_{4}$ & $1.0$ & $1.6$ & $-3.0$\\
$a_{5}$ & $1.0$ & $1.0$ & \phantom{$-$}$1.0$\\
$a_{6}$ & $3.0$ & $3.0$  & \phantom{$-$}$3.0$\\
$a_{7}$ & $1.0$  & $1.0$ & $-1.0$\\
$a_{8}$ & $5.0$ & $5.0$ & \phantom{$-$}$5.0$ \\
Model& Identity link & $\log(\lambda) = y$ & $S(t|y) = \exp[-H_{0}(t)\exp(y)]$\\
& & $t \sim \operatorname{Pois}(\lambda)$ & $H_{0}(t) = \lambda t, \lambda = 0.2$\\
& & & $t = M/(\lambda\exp(y)), M \sim \operatorname{Exp}(1)$\\
& & & $5\%$ uniform randomly censored,\\
& & & $t_{\mathrm{cens}} = \mathcal{U}(0,t_{\mathrm{event}})$\\
Train/test& $100/20$ & $100/20$ & $100/60$ \\[5pt]
Correctly selected & $6$ out of $6$ & $6$ out of $6$ & $5$ out of $6$ \\
False positives& $0$ & $0$ & $0$ \\
MSPE (normalized) & $0.0067$ & $0.045$ & see Figure \ref{coxfig1}  \\
\hline
\end{tabular*}
\end{table*}

\subsection{Large $p$}

Next we show simulation results on continuous, count and survival data
models, for $(n,p) = (100,\break 1{,}000)$. We employ an additive term as the
kernel for all models,
\begin{eqnarray}
y &=& a_{1}x_{1} + a_{2}x_{2} + a_{3}x_{3} + a_{4}x_{4} \nonumber
\\[-8pt]\\[-8pt]
&&{}+ a_{5}\sin(a_{6}
x_{5}) + a_{7}\sin(a_{8}x_{6}) + \varepsilon.\nonumber
\end{eqnarray}
The functional form for the simulation kernel is designed so that the
first four covariates express a linear relationship to the response
while the next two express nonlinear associations. Model-specific
coefficient values are displayed in Table~\ref{table:simresults}.
Methods employed to randomly generate the observed count and event time
data from the latent response kernel are also outlined in the table.
For example, the kernel captures the $\log$-mean of the Poisson
distribution used to generate count data, and it is used to generate
the survivor function that is inverted to provide event time data for
the Cox model. As in the previous simulation, all covariates are
generated from $\mathcal{U}(0,1)$.

\begin{figure*}

\includegraphics{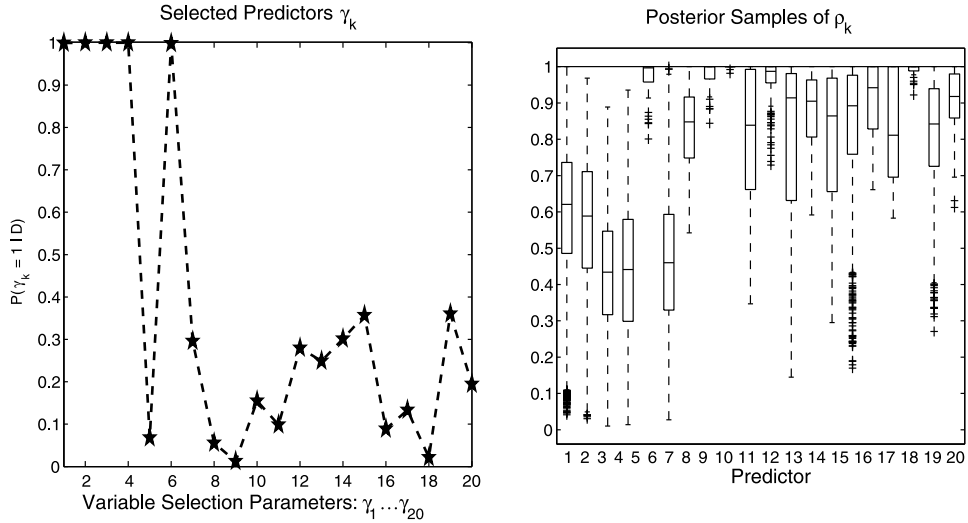}

\caption{Cox GP model with large $p$: Simulated data $(n=100, p =
1{,}000)$. Posterior distributions for $\gamma_{k} = 1$ and box plots of
posterior samples for $\rho_{k}$.} \label{coxfig2}
\vspace*{6pt}
\end{figure*}

We set the hyperparameters as described in Section~\ref{sse:hyper}. We
used MCMC scheme~1 and increased the number of total iterations, with
respect to the simpler simulation with only $p=20$, to  800,000
iterations, discarding half of them for burn-in.

\begin{figure}

\includegraphics{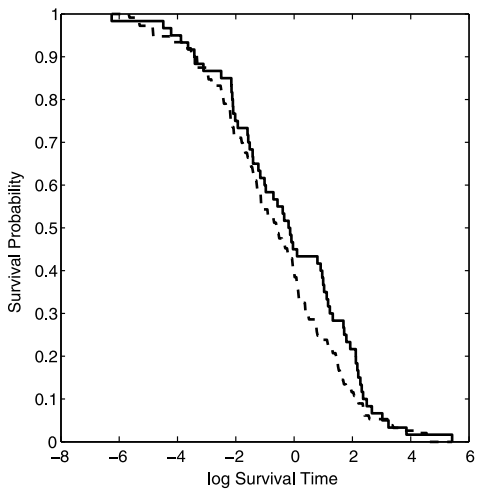}

\caption{Cox GP model with large $p$: Simulated data $(n=100, p =
1{,}000)$. Average survivor function curve on the validation set (dashed
line) compared to the Kaplan--Meier empirical estimate (solid line).}
\label{coxfig1}
\end{figure}

Results are reported in Table~\ref{table:simresults}. While the
continuous and count data GP models readily assigned high marginal
posterior probabilities to the correct covariates (figures not shown),
the Cox GP model correctly identified only $5$ of $6$ predictors; see
Figure \ref{coxfig2} for the posterior distributions of $\gamma_{k}=1$
and the box plots for the posterior samples of $\rho_{k}$ for this
model (for readability, only the first $20$ covariates are displayed).
The predictive power for the continuous and count data models was
assessed by normalizing the mean squared prediction error (MSPE) with
the variance of the test set. Excellent results were achieved in our
simulations. For the Cox GP model, the averaged survivor function
estimated on the test set is shown in Figure \ref{coxfig1}, where we
observe a tight fit between the estimated curve and the Kaplan--Meier
empirical estimate constructed from the same test data.

Though for the Cox model we only report results obtained using the
partial likelihood formulation, we conducted the same simulation study
with the model based on the full likelihood of \citet{kalb1978}. The
partial likelihood model formulation produced more consistent results
across multiple chains, with the same data, and was able to detect much
weaker signals. The \citet{kalb1978} model did, however, produce lower
posterior values near $0$ for nonselected covariates, unlike the
partial likelihood formulation, which shows values typically from
10--40\%, pointing to a potential bias toward false positives.

Additional simulations, including larger sample si\-zes cases, are
reported in \citet{terrance2010}.

\begin{table*}[t!]
\caption{Efficiency comparison of GP MCMC methods}
\label{table:simMCMC}
\begin{tabular}{@{}lccc@{}}
\hline
&\multicolumn{2}{c}{\textbf{MCMC scheme 2}} &\textbf{MCMC scheme 1}\\[-5pt]
&\multicolumn{3}{c@{}}{\hrulefill}\\
& \textbf{Adaptive} & \textbf{Nonadaptive} &  \\
\hline
Iterations (computation) & $5{,}000$& $5{,}000$ & $500{,}000$ \\
Autocorrelation time & & &\\
$\rho_{6}$ & $310$ & $82$ & $441$ \\
$\rho_{8}$ & $59$ & $35$ & $121$ \\[5pt]
Computation& & & \\
CPU-time (sec) & $980$ & $4{,}956$ & $10{,}224$ \\
\hline
\end{tabular}
\end{table*}

\subsection{Comparison of MCMC Methods}\label{mcmcmethods}

We compare the 2 MCMC schemes previously described for posterior
inference on $(\bolds{\gamma},\bolds{\rho})$ on the basis of sampling
and computational efficiency.  We use the univariate regression
simulation kernel
\begin{eqnarray*}
y &=& x_{1} + 0.8x_{2} + 1.3x_{3} + \sin(x_{4}) + \sin(3x_{5}) \\
&&{}+ \sin(5x_{6}) + (1.5x_{7})(1.5x_{8}) +  \varepsilon,
\end{eqnarray*}
with $\varepsilon \sim \mathcal{N}(0,\sigma^{2})$ and $\sigma =0.05$.
We utilize $1{,}000$ covariates with all but the first $8$ defined as
nuisance.  We use a training and a validation set of $100$ observations
each.

The two schemes differ in the way they update
$(\bolds{\gamma},\bolds{\rho})$. While scheme 1 samples either one or
two positions in the model space on each iteration, scheme~2 samples
$(\gamma_{k},\rho_{k})$ for each of the $p$ covariates. Because of this
a good ``rule-of-thumb'' should employ a number of iterations for
scheme 1 which is roughly $p$ times the number of iterations employed
for scheme~2. The use of the \textit{Keep} move in scheme~1, however,
reduces the need of scaling the number of iterations by exactly $p$,
since all $\rho_k$'s are sampled at each iteration. In our simulations
we found stable convergence under moderate correlation among covariates
for scheme~2 in $5{,}000$ iterations and for scheme 1 in $500{,}000$
iterations. For both schemes, we discarded half of the iterations as
burn-in. The CPU run times we report in Table \ref{table:simMCMC} are
based on utilization of Matlab with a 2.4~GHz Quad Core (Q6600) PC with
4~GB of RAM running 64-bit Windows XP.

We compared sampling efficiency looking at autocorrelation for selected
$\rho_{k}$.  The autocorrelation time is defined as one plus twice the
sum of the autocorrelations at all lags and serves as a measure of the
relative dependence for MCMC samples.  We used the number of MCMC
iterations divided by this factor as an ``effective sample size.''  We
followed a procedure outlined by \citet{neal2000} and ran first scheme~2 for $1{,}000$ iterations, to obtain a state near the posterior
distribution. We then employed this state to initiate a chain for each
of the two schemes. We ran scheme~2 for an additional $2{,}000$
iterations and scheme 1 for $200{,}000$ (using the last $2{,}000$ draws
for each of the target $\rho_{k}$ for final comparison). For scheme~2
we used both the adaptive and nonadaptive versions. Table
\ref{table:simMCMC} reports results for $\rho_{8}$, aligned to a
covariate expressing a linear interaction, and for $\rho_{6}$, for a
highly nonlinear interaction. We observe that both versions of scheme~2
express notable improvements in computational efficiency as compared to
scheme 1. We note, however, that the adaptive scheme method produces
draws of higher autocorrelation than the nonadaptive method.

\begin{figure*}

\includegraphics{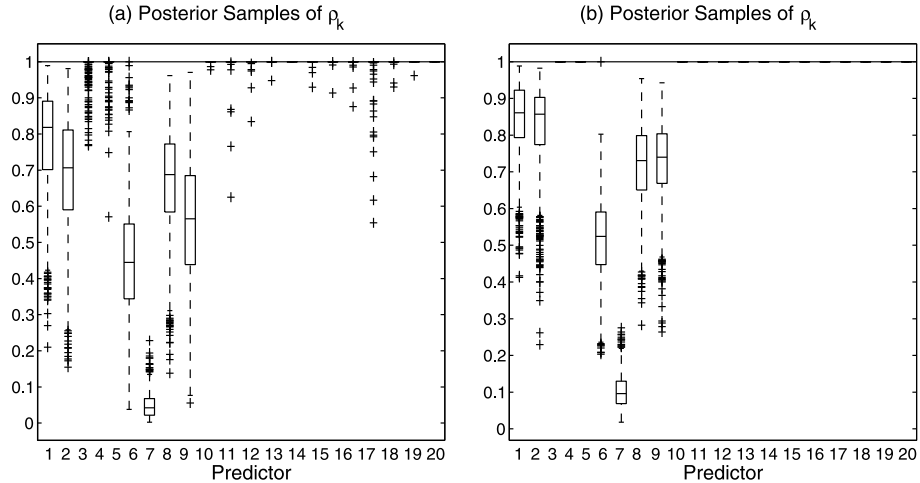}

 \caption{Prior Sensitivity for $\rho_{k}|\gamma_{k}=1 \sim
\operatorname{Beta}(a,b)$: Box plots of posterior samples for
$\rho_{k}$ for $(a,b)=(0.5,0.5)$---plot (\textup{a})---and
$(a,b)=(2.0,2.0)$---plot (\textup{b}).}
\label{rhoprior}
\vspace*{-3pt}
\end{figure*}

\begin{table}
\caption{Prior sensitivity for $\rho_{k}|\gamma_{k}=1 \sim
\operatorname{Beta}(a,b)$. Results are reported as (number of false
negatives)${}/{}$(normalized MSPE)} \label{tab:rhosens}
\begin{tabular*}{\columnwidth}{@{\extracolsep{\fill}}lccc@{}}
\hline
$\bolds{b\backslash a}$ & $\bolds{0.5}$ & $\bolds{1.0}$ & $\bolds{2.0}$ \\
\hline
\textbf{0.5} & $2/0.18$ & $2/0.15$ & $2/0.18$ \\
\textbf{1.0} & $1/0.14$ & $1/0.16$ & $2/0.18$\\
\textbf{2.0} & $1/0.15$ & $2/0.16$ & $2/0.17$ \\
\hline
\end{tabular*}
\end{table}

\subsection{Sensitivity Analysis} \label{sensitivity}

We begin with a sensitivity analysis on the prior~for
$\rho_{k}|\gamma_{k}=1$.  Table \ref{tab:rhosens} shows results under
a full~fac\-torial combination for hyperparameters $(a,b)$ of a~Beta~prior
construction, where we recall $\operatorname{Beta}(1,1) \equiv
\mathcal{U}(0,1)$. Results were obtained with the univariate regression
simulation kernel
\begin{eqnarray*}
y &=& x_{1} + x_{2} + \sin(1.5x_{3})\sin(1.5x_{4}) + \sin(3x_{5}) \\
&&{}+ \sin(3x_{6})+ (1.5x_{7})(1.5x_{8}) +\varepsilon,
\end{eqnarray*}
with $\varepsilon \sim \mathcal{N}(0,\sigma^{2})$ and where we employed
a higher error variance of $\sigma =0.28$. As before, we employ
$1{,}000$ covariates with all but the first $8$ defined as nuisance.  A
training sample of $110$ was simulated, along with a test set of $100$
observations.  We employed the adaptive scheme~2, with $5{,}000$
iterations, half discarded as burn-in.

Figure \ref{rhoprior} shows box plots of posterior samples for $\rho_k$
for two symmetric alternatives, $1\dvtx (a,b) = (0.5,0.5)$ (U-shaped)
and $2\dvtx (a,b) = (2.0,2.0)$ (symmetric unimodal). For scenario 2 we
observe a reduction in posterior jitter on nuisance covariates and a
stabilization of posterior sampling for associated covariates, but also
a greater tendency to exclude $x_{3}, x_{4}$.  One would expect the
differences in posterior sampling behavior across prior hyperparameter
values to decline as the sample size increases. Table \ref{tab:rhosens}
displays the number of nonselected true variables (false negatives),
out of 8, along with the normalized MSPEs for all scenarios. There were
no false positives to report across all hyperparameter settings.
Overall, results are similar across the chosen settings for $(a,b)$,
with slightly better performances for $a<1$ and $b\geq 1$,
corresponding to strictly decreasing shapes that aid selection by
pushing more mass away from 1, increasing the prior probability of the
good variables to be selected, especially in the presence of a large
number of noisy variables.

Next we imposed a Beta distribution on the hyperparameter $\alpha$ of
the priors $\gamma_{k} \sim \operatorname{Bernoulli}(\alpha)$ for
covariate inclusion. We follow \citet{brown1998b} to specify a vague
prior by setting the mean of the Beta prior to $0.025$, reflecting
a~prior expectation for model sparsity, and the sum of the two parameters
of the distribution to $2$. We ran the same univariate regression
simulation kernel as above with the hyperparameter settings for the
Beta prior on $\rho_{k}$ equal to $(1,1)$ and obtained the same
selection results as in the case of $\alpha$ fixed and a  slightly
lower normalized MSPE of $0.14$.

\begin{table*}
\caption{Ozone data: Results}\label{tab:ozoneResults}
\begin{tabular}{@{}lccc@{}}
\hline
\textbf{Prior on} $\bolds{g}$ & $\bolds{\mathcal{M}}_{\bolds{\gamma}}$ & $\mathbf{p}_{\bolds{\gamma}}$ & \textbf{RMSPE}  \\
\hline
Local empirical Bayes & $X_{5}, X_{6}, X_{7}, X_{6}^{2}, X_{7}^{2}, X_{3}X_{5}$ & 6 &
4.5\\[2pt]
Hyper-$g$ ($a=4$) & $X_{5}, X_{6}, X_{7}, X_{6}^{2}, X_{7}^{2}, X_{3}X_{5}$ & 6 & 4.5 \\[2pt]
Fixed (BIC) & $X_{5}, X_{6}, X_{7}, X_{6}^{2}, X_{7}^{2}, X_{3}X_{5}$ & $6$ & 4.5\\[2pt]
\citet{brown2002} &  $X_{1}X_{6}, X_{1}X_{7}, X_{6}X_{7}, X_{1}^{2}, X_{3}^{2}, X_{7}^{2}$ & 6 & 4.5\\[2pt]
GP model & $X_{3}, X_{6}, X_{7}$ & 3 & 3.7 \\
\hline
\end{tabular}
\end{table*}

Last, we explored performances with respect to correlation among the
predictors. We utilized the same kernel as above with $8$ true
predictors from which to construct the response. We then induced a
$70\%$ correlation among $20$ randomly chosen nuisance covariates and
the true predictor $x_6$. We found $2$ false negatives and $1$ false
positive, which demonstrates a relative selection robustness under
correlation.  We did observe a significant decline in normalized MSPE,
however, to $0.33$, as compared to previous runs.

\section{Benchmark Data Applications}\label{applications}

We now present results on two data sets often used in the literature as
benchmarks. For both analyses we performed inference by using the
MCMC---scheme~2, with $5{,}000$ iterations and half discarded as
burn-in.

\subsection{Ozone data}

We start by revisiting the ozone data, first analyzed for variable
selection by \citet{breiman1985} and more recently by
\citet{liang2007}. This data set supplies integer counts for the
maximum number of ozone particles per one million particles of air near
Los Angeles for $n = 330$ days and includes an associated set of $8$
meteorological predictors. We held out a randomly chosen set of $165$
observations for validation.

\begin{figure*}

\includegraphics{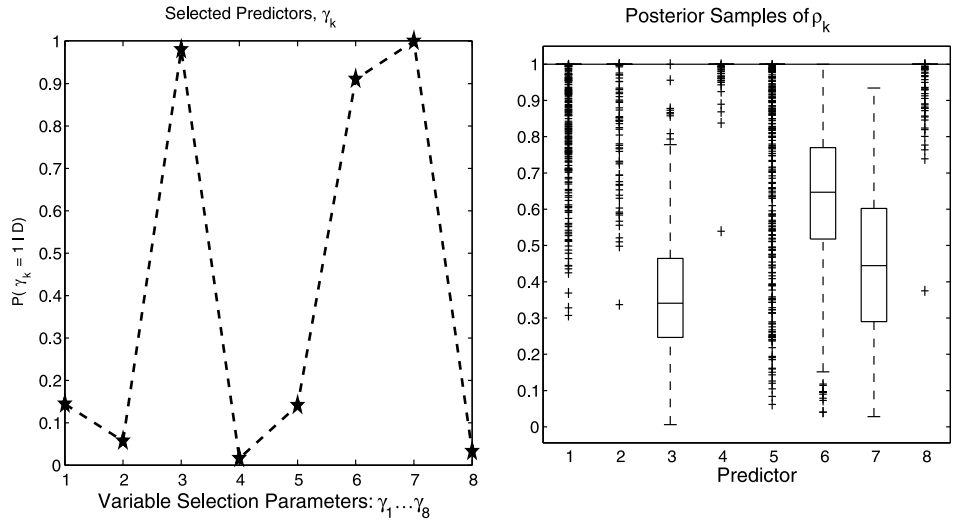}

\caption{Ozone data: Posterior distributions for $\gamma_{k} = 1$
and box plots of posterior samples for $\rho_{k}$.} \label{ozone}
\vspace*{-2pt}
\end{figure*}

\begin{figure*}

\includegraphics{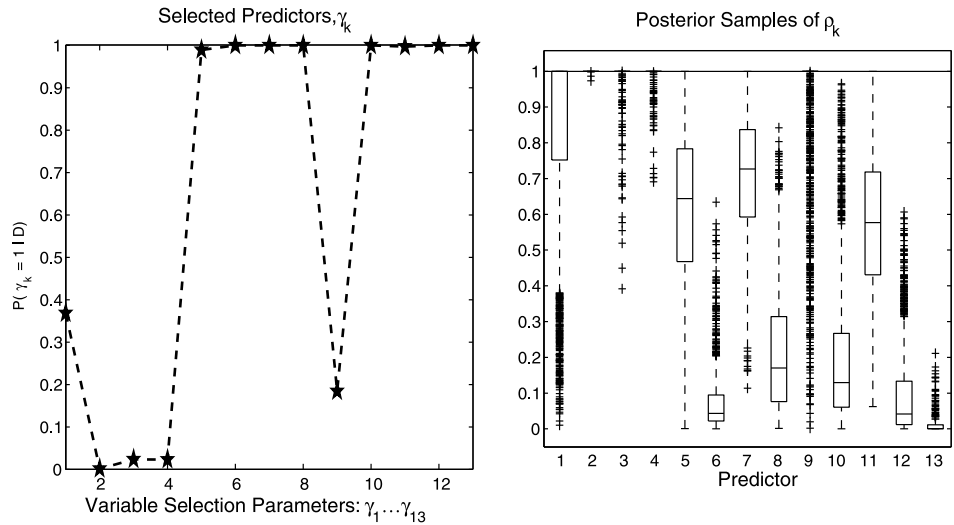}

\caption{Boston housing data: Posterior distributions for $\gamma_{k} =
1$ and box plots of posterior samples for $\rho_{k}$.}\label{BH}
\vspace*{-2pt}
\end{figure*}

\citet{liang2007} use a linear regression model including all linear
and quadratic terms for a total of $p = 44$ covariates. They achieve
variable selection by imposing a mixture prior on the vector
$\bolds{\beta}$ of regression coefficients and specifying a $g$-prior
of the type $\bolds{\beta}_{\bolds{\gamma}}|\phi \sim
\mathcal{N}(\mathbf{0},\frac{g}{\phi}(\mathbf{X}_{\bolds{\gamma}}^{T}
\mathbf{X}_{\bolds{\gamma}})^{-1})$. Their results are reported in
Table \ref{tab:ozoneResults} with various formulations for $g$.  In
particular, the local empirical Bayes method offers a model-dependent
maximizer of the marginal likelihood on $g$, while the hyper-$g$
formulation with $a = 4$ is one member of a continuous set of
hyper-prior distributions on the shrinkage factor, $g/(1+g)\sim
\operatorname{Beta}(1,a/2-1)$. Since the design matrix expresses a high
condition number, a situation that can at times induce poor results
with $g$-priors, we additionally applied the method of
\citet{brown2002} who used a mixture prior of the type
$\bolds{\beta}_{\bolds{\gamma}} \sim
\mathcal{N}(\mathbf{0},c\mathbf{I})$. Results shown in Table
\ref{tab:ozoneResults} were obtained from the Matlab code made
available by the authors.

Though previous variable selection work on the ozone data all choose a
Gaussian likelihood, a more precise approach employs a discrete Poisson
or negative binomial formulation on data with low count values, or a
log-normal approximation where counts are high.  With a maximum value
of $38$ and a mean of $11$ we chose to model the data with the
negative-binomial count data model (\ref{nbin}). We used the same
hyperparameter settings as in our simulation study. Results are shown
in Figure \ref{ozone}. By selecting, for example, the best~3 variables,
we achieve a notable decrease in the root-MSPE as compared to the
linear models. Also, by allowing an a priori unspecified functional
form for how covariates relate to the response, we end up selecting a
much more parsimonious model, although, of course, we lose in
interpretability of the selected terms, with respect to linear
formulations that specifically include linear, quadratic and
interactions terms in the model.

\subsection{Boston Housing data}

Next we utilize the Boston Housing data set, also analyzed by
\citet{breiman1985}, who used an additive model and employed an
algorithm to empirically determine the functional relationship for each
predictor. This data set relates $p = 13$ predictors to the median
value of owner-occupied homes in each of $n = 506$ census tracts in the
Boston metro\-politan area. As with the previous data set, we held out a
random set of $250$ observations to assess prediction.

We employed the continuous data model (\ref{uni}) with the same
hyperparameter settings as in our simulations.  The four predictors
chosen by \citet{breiman1985}, $(x_{6}, x_{10}, x_{11}, x_{13}),$ had
all marginal posterior probability of inclusion greater than 0.9 in our
model. Other variables with high marginal posterior probability were
$(x_{5}, x_7, x_{8}, x_{12})$. The adaptability of the GP response
surface is illustrated with closer examination of covariate $x_{5}$,
which measures the level of nitrogen oxide (NOX), a pollutant emitted
by cars and factories.  At low levels, indicating proximity to jobs,
$x_{5}$ presents a positive association to the response, and at high
levels, indicating overly industrialized areas, a negative association.
This inverted parabolic association over the covariate range probably
drove its exclusion in the model of \citet{breiman1985}. The GP
formulation is, however, able to capture this strong nonlinear
relationship as is noted in Figure \ref{BH}. By using only the subset
of the best eight predictors, we achieved a~normalized MSE of $0.1$ and
a prediction $R^{2}$ of $0.9$, very close to the value of $0.89$
reported by \citet{breiman1985} on the training data.

We also employed the Matern covariance construction (\ref{matern}),
which we recall employs an explicit smoothing parameter, $\nu \in
[0,\infty)$. While selection results were roughly similar, the
prediction results for the Matern model were significantly worse than
the exponential model, with a normalized MSPE of $0.16$, probably due
to overfitting.  It is worth noticing that the more complex form for
the Bessel function increases the CPU computation time by a factor of
5--10 under the Matern covariance as compared to the exponential
construction.

\begin{figure*}

\includegraphics{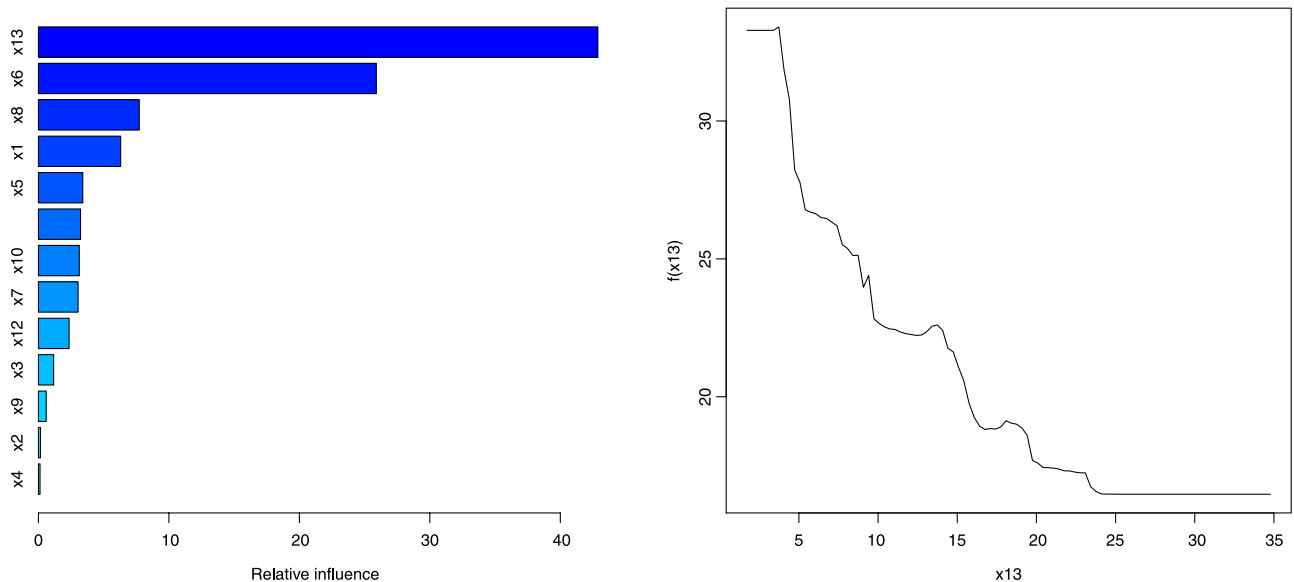}

\caption{Boston housing data: GBM covariate analysis. Left-hand chart
provides variables importance, normalized to sum up to 100. Right-hand
plot enumerates partial association of $x_{13}$ to the response.}
\label{GBM}
\end{figure*}

For comparison, we looked at GBMs. We used version 3.1 of the
\textbf{gbm} package for the R software environment. We utilized the
same training and validation data as above.  After experimentation and
use of $10$-fold cross-validation, we chose a small value for the input
regularization parameter, $\nu =0.0005$, to provide a smoother fit that
prevents overfitting.  Larger values of $\nu$ resulted in higher
prediction errors. The GBM was run for $50{,}000$ iterations to achieve
minimum fit error. The result provided a~normalized MSPE of $0.13$ on
the test set, similar to, though slightly higher than, the GP result.
The left-hand chart of Figure \ref{GBM} displays the relative covariate
importance. Higher values correspond to $(x_{13}, x_{6}, x_{8})$, and
agree with our GP results.  A number of other covariates show similar
importance values to one another, though lower than these top $3$,
making it unclear as to whether they are truly related or nuisance
covariates. Similar conclusions are reported by other authors.  For
example, \citet{tokdar2010} analyze a subset of the same data set with
a Bayesian density regression model based on logistic Gaussian
processes and subspace projections and found $(x_{13},x_6)$ as the most
influential predictors, with a number of others having a mild influence
as well. The right-hand plot supplies a partial dependence plot
obtained by the GBM for variable $x_{13}$ by averaging over the
associations for the other covariates.  We note that the nonlinear
association is not constrained to be smooth under~GBM.

\section{Discussion} \label{discussion}

In this paper we have presented a unified  modeling approach via
Gaussian processes that extends to data from the exponential dispersion
family and to survival data.  Such model formulation allows for
nonlinear associations of the predictors to the response.  We have
considered, in particular, continuous, categorical and count responses
and survival data. Next we have addressed the important problem of
selecting variables from a set of possible predictors and have put
forward a general framework that employs Bayesian variable selection
methods and mixture priors for the selection of the predictors. We have
investigated strategies for posterior inference and have demonstrated
performances on simulated and benchmark data. GP models provide a~%
parsimonious approach to model formulation with a great degree of
freedom for the data to define the fit. Our results, in particular,
have shown that GP models can achieve good prediction performances
without the requirement of prespecifying higher order and nonlinear
additive functions of the predictors. The benchmark data applications
have shown that a GP formulation may be appropriate in cases of
\textit{heterogeneous} covariates, where the inability to employ an
obvious transformation would require\break higher order polynomial terms in
an additive linear fashion, or even in the case of a homogeneous
covariate space where the transformation overly reduces structure in
the data.  Our simulation results have further highlighted the ability
of the GP formulation to manage data sets with $p \gg n$.

A challenge in the use of variable selection methods in the GP
framework is to manage the numerical instability in the construction of
the GP covariance matrix. In the \hyperref[app]{Appendix} we describe a
projection method to reduce the effective dimension of this matrix.
Another practical limitation of the models we have described is the
difficulty to use them with qualitative predictors.  \citet{qian2008}
provide a modification of the GP covariance kernel that allows for
nominal qualitative predictors consisting of any number of levels.  In
particular, the authors model the covariance structure under a mixture
of qualitative and quantitative predictors by employing a
multiplicative factor against the usual GP kernel for each qualitative
predictor to capture the by-level categorical effects.

Some generalization of the methods we have presented are possible. For
example, as with GLM models, we may employ an additional set of
variance inflation parameters in a similar construction to
\citet{neal1999} and others to allow for heavier tailed distributions
while maintaining the conjugate framework.

\appendix
\section*{Appendix: Computational Aspects}\label{app}

We focus on the exponential form (\ref{covmatrix}) and introduce an
efficient computational algorithm to generate ${\bf C}$.  We also
review a method of \citet{banerjee2008} to approximate the inverse
matrix that employs a random subset of observations and provide a
pseudo-code.

\subsection{Generating the Covariance Matrix $\mathbf{C}$}

Let us begin with the quadratic expression, $\mathbf{G}=\{g_{i,j}\}$ in
(\ref{covmatrix}).  We rewrite $g_{i,j} =
A_{i,j}'[-\log(\bolds{\rho})]$ with $A_{i,j}$ constructed as a $p
\times 1$ vector of term-by-term squared differences,
$(x_{ik}-x_{jk})^{2}, k = 1, \ldots, p$. We may directly employ the $p
\times 1$ vector, $\bolds{\rho}$, as $\mathbf{P}$ is diagonal. As a
first step, we may then directly compute $\mathbf{G} =
\mathbf{A}[-\log(\bolds{\rho})]$, where $\mathbf{A}$ is $n\times n
\times p$.  We are, however, able to reduce the more complex structure
of $\mathbf{A}$ to a two dimensional matrix form by simply stacking
each $\{i,j\}$ row of dimension $1\times p$ under each other such that
our revised structure, $\mathbf{A}^{*}$, is of dimension $n^2\times p$
and the computation, $\mathbf{G} =
\mathbf{A}^{*}[-\log(\bolds{\rho})]$, reduces to a series of inner
products.  Next, we note that $\log(\rho_{k}) = 0$ for $\rho_{k}=1$. So
we may reduce the dimension for each of the $n^{2}$ inner products by
reducing the dimension of $\bolds{\rho}$ to the $p_{\bolds{\gamma}} <
p$ nontrivial covariates.  We may further improve efficiency by
recognizing that since our resultant covariance matrix, $\mathbf{C}$,
is symmetric positive definite, we need only compute the inner products
for a reduced set of unique terms (by removing redundant rows from
$\mathbf{A}^{*}$) and then ``re-inflate'' the result to a vector of the
correct length.  Finally, we exponentiate this vector, multiply the
nonlinear weight ($1/\lambda_{z}$), add the affine intercept term,
($1/\lambda_{a}$), and then reshape this vector into the resulting
$n\times n$ matrix, $\mathbf{C}$.  The resulting improvement in
computational efficiency at $n = 100$ from the naive approach that
employs double loops of inner products is on the order of $500$ times.

Our MCMC scheme~2 proposes a change to $\rho_{k} \in \bolds{\rho}$,
one-at-a-time, conditionally on $\bolds{\rho}_{-k}$ and the other
sampled parameters.  Changing a single $\rho_{k}$ requires updating
only one column of the inner product computation of $\mathbf{A}^{*}$
and $[-\log(\bolds{\rho})]$.  Rather than conducting an entire
recomputation for $\mathbf{C}$, we multiply the $k$th column of
$\mathbf{A}^{*}$ (with number of rows reduced to only unique terms in
$\mathbf{C}$) by
$\log(\frac{\rho_{k,\mathrm{prop}}}{\rho_{k,\mathrm{old}}})$, where
``prop'' means the proposed value for $\rho_{k}$.  This result is next
exponentiated (to a covariance kernel), re-inflated and shaped into an
$n\times n$ matrix, $\bolds{\Delta}$.  We then take the current value
less the affine term, $\mathbf{C}_{\mathrm{old}} -
\frac{1}{\lambda_{a}}\mathbf{J}_{n}$, and multiply by
$\bolds{\Delta}$, term-by-term, and add back the affine term to
achieve the new covariance matrix associated to the proposed value for
$\rho_{k}$. So we may devise an algorithm to update an existing
covariance matrix, C, rather than conducting an entire recomputation.
At $p = 1{,}000$ with $6$ nontrivial covariates and $n = 100$, this
algorithm further reduces the computation time over recomputing the
full covariance by a factor of $2$.  This efficiency grows nonlinearly
with the number of nontrivial covariates.

\subsection{Projection Method for Large $n$}\label{sse:proj}

In order to ease the computations, we have also adapted a dimension
reduction method proposed by \citet{banerjee2008} for spatial data. The
method achieves a reduced-dimension computation of the inverse of the
full ($n \times n$) covariance matrix. It can also help with the
accuracy and stability of the posterior computations when working with
possibly ill-conditioned GP covariance matrices, particularly for large
$n$. To begin, randomly choose $m<n$ points (knots), sampled within
fixed intervals on a grid to ensure relatively uniform coverage, and
label these $m$ points $\mathbf{z}^{*}$.  Then define $\mathbf{z}_{m
\rightarrow n}$ as the orthogonal projection of $\mathbf{z}$ onto the
lower dimensional space spanned by $\mathbf{z}^{*}$, computed as the
conditional expectation
\[
\mathbf{z}_{m \rightarrow n} = \mathbb{E}(\mathbf{z}|\mathbf{z}^{*}) =
\mathbf{C}_{(\mathbf{z}^{*},\mathbf{z})}'\mathbf{C}_{(\mathbf{z}^{*},\mathbf{z}^{*})}^{-1}\mathbf{z}^{*}.
\]
We use the univariate regression framework in (\ref{uni}) to illustrate
the dimension reduction from constructing the \textit{projection} model
using $\mathbf{z}_{m \rightarrow n}$ in place of
$\mathbf{z}(\mathbf{x})$.  Recast the model from (\ref{uni}) to
\[
\mathbf{y} = \mathbf{z}_{m \rightarrow n} + \bolds{\varepsilon}=
\mathbf{C}_{(\mathbf{z}^{*},\mathbf{z})}'\mathbf{C}_{(\mathbf{z}^{*},\mathbf{z}^{*})}^{-1}\mathbf{z}^{*}
+\bolds{\varepsilon},
\]
where $\varepsilon_{i} \sim \mathcal{N}(0,\frac{1}{r})$. Then derive
$\bolds{\Lambda}_{n} = \operatorname{Cov}(\mathbf{y})=\break
\frac{1}{r}\mathbb{I}_{n}+\mathbf{C}_{(\mathbf{z}^{*},\mathbf{z})}'\mathbf{C}_{(\mathbf{z}^{*},\mathbf{z}^{*})}^{-1}\mathbf{C}_{(\mathbf{z}^{*},\mathbf{z})}$.
Finally,\vspace*{2pt} employ the\break Woodbury matrix identity to transform the inverse
computation, $\bolds{\Lambda}_{n}^{-1} = r\mathbb{I} -
r^{2}\mathbf{C}_{(\mathbf{z}^{*},
\mathbf{z})}'[\mathbf{C}_{(\mathbf{z}^{*},\mathbf{z}^{*})}+\break
r\mathbf{C}_{(\mathbf{z}^{*},\mathbf{z})}\cdot\mathbf{C}_{(\mathbf{z}^{*},
\mathbf{z})}']^{-1}\mathbf{C}_{(\mathbf{z}^{*}, \mathbf{z})}$, where
the quantity inside the square brackets, now being inverted, is $m
\times m$, supplying the dimension reduction for inverse computation we
seek. We note that, in the absence of the projection method, a large
jitter term would be required to invert the GP covariance matrix,
trading accuracy for stability.  Though the projection method
approximates a higher dimensional covariance matrix in a lower
dimensional projection, we yet improve performance and avoid the
accuracy/sta\-bility trade-off.  We do, however, expect to use more
iterations for MCMC convergence when employing a relatively lower
projection ratio.

All results shown in this paper were obtained with $m/n = 0.35$, for
simulated data, and with $m/n = 0.25$, for the benchmark applications,
where we enhanced computation stability in the presence of the high
condition number for the design matrix. We have also employed the
Cholesky decomposition, in a similar fashion as in \citet{neal1999}, in
lieu of directly computing the resulting $m \times m$ inverse.

\subsection{Pseudo-code}

Procedure to Compute, $\mathbf{C} = \frac{1}{\lambda_{a}}\mathbf{J}_{n}
+ \frac{1}{\lambda_{z}}\exp(-\mathbf{G})$:

\begin{tabbing}
\textbf{Input}\=: da\=ta matric\=es;\\
\hphantom{\textbf{Input}\ \ }$(\mathbf{X}_{1}, \mathbf{X}_{2})$ of dimension $(n_{1},n_{2})\times p$\\
\textbf{Output}: function, \textbf{ [$A^{*}, I_{\mathrm{full}}] ={}$difference($\mathbf{X}_{1},\mathbf{X}_{2}$)}\\
\% $A^{*}$ is matrix of squared $L_{2}$ distances\\
\hphantom{\%\ }for 2 data matrices of $p$ columns\\
\% $A^{*}$ size, $\ell \times p,~\ell \leq n_{1}n_{2}$: only unique entries\\
\% $I_{\mathrm{full}}$ re-inflates $A^{*}$ with duplicate entries\\
\% Key point: Compute $A^{*}$, once,\\
\hphantom{\%\ }and re-use in GP posterior computations\\
\% Set counter to stack all $(i,j)$ obs\\
\hphantom{\%\ }from $\mathbf{X}_{1}, \mathbf{X}_{2}$ in vectorized construction\\
\> count${}={}$1; \\
\% Compute squared distances \\
\> \textbf{FOR} $i = 1$ to $n_{1}$\\
\> \> \textbf{FOR} $j = 1$ to $n_{2}$\\
\> \> \> $A_{\mathrm{full}}^{*}$(count,:) = $(x_{1,i} - x_{2,j})^{2}$;\\
\> \> \> count${}={}$count${}+{}$1; \\
\> \> \textbf{END} \\
\> \textbf{END}\\
\% Reduce $A_{\mathrm{full}}^{*}$ to $A^{*}$\\
\> $[A^{*}, I_{\mathrm{full}}] = \mbox{unique}(A_{\mathrm{full}}^{*},\mbox{by row})$;\\
\textbf{END FUNCTION}\\
\\
\textbf{Input}: Data${}={}$($A^{*}, I_{\mathrm{full}})$,  $\bolds{\Theta} = (\bolds{\rho}, \lambda_{a}, \lambda_{z})$\\
\textbf{Output}: function, $[\mathbf{C}]$${}={}$C($A^{*}, I_{\mathrm{full}}, \bolds{\Theta}$)\\
\% An $n_{1}\times n_{2}$ GP covariance matrix\\
\% Only compute inner product\\
\hphantom{\%\ }for column $k$ where $\rho_{k} < 1$\\
\>$\mathit{sel}_{\rho} = \{\rho_{k} < 1\}$;\\
\>$\bolds{\rho} = \bolds{\rho}(\mathit{sel}_{\rho})$\\
\>$A^{*} = A^{*}(:,\mathit{sel}_{\rho})$;\\
\% Compute vector of unique values for $\mathbf{C}$\\
\>$-G_{\mathrm{vec}} = A^{*}[\log(\bolds{\rho})]'$;\\
\>$C_{\mathrm{vec}} = \frac{1}{\lambda_{a}} + \frac{1}{\lambda_{z}}\exp(-G_{\mathrm{vec}})$;\\
\% Re-inflate $C_{\mathrm{vec}}$ to include duplicate values\\
\>$C_{\mathrm{vec}} = C_{\mathrm{vec}}(I_{\mathrm{full}})$;\\
\% Snap $C_{\mathrm{vec}}$ into matrix form, $C$\\
\>$C = \mbox{reshape}(C_{\mathrm{vec}},n_{2},n_{1})'$;\\
\textbf{END FUNCTION}\\
\\
\textbf{Input}: Previous covariance${}={}$$C_{\mathrm{old}}$;\\
\hphantom{\textbf{Input}:\ }Data${}={}$($A^{*}, I_{\mathrm{full}})$; Position changed${}=k$, \\
\hphantom{\textbf{Input}:\ }Parameters${}=(\rho_{k,\mathrm{new}},\rho_{k,\mathrm{old}})$, Intercept${}=\lambda_{a}$ \\
\textbf{Output}:$[\mathbf{C}_{\mathrm{new}}] = C_{\mathrm{partial}}(C_{\mathrm{old}},A^{*},I_{\mathrm{full}},k,\lambda_{a})$\\
\% Compose new covariance matrix, $\mathbf{C}_{\mathrm{new}}$,\\
\hphantom{\%} from old, $\mathbf{C}_{\mathrm{old}}$\\
\% Compute inner products only for row $k$ of $A^{*}$\\
\% Produce matrix of multiplicative differences\\
\hphantom{\%}  from old to new\\
\> $-\bolds{\Delta} \mathbf{G}_{\mathrm{vec}} = \mathbf{A}^{*}(:,k) \times \log(\frac{\rho_{k,\mathrm{new}}}{\rho_{k,\mathrm{old}}})$;\\
\% Re-inflate $\exp(-\bolds{\Delta} \mathbf{G}_{\mathrm{vec}})$\\
\> $\exp(-\bolds{\Delta} \mathbf{G}_{\mathrm{vec}}) = \exp(-\bolds{\Delta} \mathbf{G}_{\mathrm{vec}})(I_{\mathrm{full}})$;\\
\% Re-shape $-\bolds{\Delta} \mathbf{G}_{\mathrm{vec}}$ to matrix, $\bolds{\Delta}$\\
\> $\bolds{\Delta} = \mbox{reshape}(\exp[-\bolds{\Delta} \mathbf{G}_{\mathrm{vec}}],n_{2},n_{1})'$;\\
\% Compute $C_{\mathrm{new}}$ \\
\> $\mathbf{C}_{\mathrm{new}} = \frac{1}{\lambda_{a}}{\bf J}_n + (\mathbf{C}_{\mathrm{old}} - \frac{1}{\lambda_{a}}{\bf J}_n)\bigodot\bolds{\Delta}$;\\
\textbf{END FUNCTION}
\end{tabbing}
Procedure to Compute Inverse of  $\bolds{\Lambda}_{\mathbf{n}} =
\frac{1}{r}\mathbb{I}_{n} + \mathbf{C}$:
\begin{tabbing}
\textbf{Input}\=: Nu\=mber \=of sub-sample${}=m$, Data${}=\mathbf{X}$,\\
\hphantom{\textbf{Input}:\ }Error precision${}=r$ \\
\hphantom{\textbf{Input}:\ }Covariance parameters =  $\bolds{\Theta} = (\bolds{\rho}, \lambda_{a}, \lambda_{z})$\\
\textbf{Output}: $\bolds{\Lambda}_{\mathbf{n}}^{-1}$\\
\% Randomly select $m<n$ observations\\
\hphantom{\%\ }on which to project $n \times 1, z(x)$\\
\> $\mathit{ind} ={}$random.permutations.latin.hypercube$(n)$;\\
\> \% space-filling \\
\> $\mathbf{X}_{m} = \mathbf{X}(ind(1\dvtx m),:)$;\\
\% Compute squared distances, $A_{m}^{*}, A^{*}$\\
\> $[A_{m}^{*}, I_{m,\mathrm{full}}] = \mbox{\small{difference}}(\mathbf{X}_{m},\mathbf{X}_{m})$; \% $m \times n$\\
\> $[A^{*}, I_{\mathrm{full}}] = \mbox{\small{difference}}(\mathbf{X}_{m},\mathbf{X})$; \% $n \times n$\\
\% Compose associated covariance matrices \\
\> $C_{(m,m)} = C(A_{m}^{*},I_{m,\mathrm{full}},\bolds{\Theta)}$;\\
\> $C_{(m,n)} = C(A^{*},I_{\mathrm{full}},\bolds{\Theta})$;\\
\% Compute $\bolds{\Lambda}_{\mathbf{n}}$\\
\> $\bolds{\Lambda}_{\mathbf{n}} = \frac{1}{r}\mathbb{I}_{n} +  C_{(m,n)}'C_{(m,m)}^{-1}C_{(m,n)}$;\\
\% Compute $\bolds{\Lambda}_{\mathbf{n}}^{-1}$ employing\\
\hphantom{\%\ }term-by-term multiplication\\
\> $\bolds{\Lambda}_{\mathbf{n}}^{-1}= r\mathbb{I}_{n} - r^{2}C'_{(m,n)}[C_{(m,m)}$ \\
\> \hphantom{$\bolds{\Lambda}_{\mathbf{n}}^{-1}={}$}$+ rC_{(m,n)}C_{(m,n)}']^{-1}C_{(m,n)}$;\\
\textbf{END}
\end{tabbing}

\section*{Acknowledgments}
Marina Vannucci supported in part by NIH-\break NHGRI Grant R01-HG003319 and
by NSF Grant DMS-10-07871. Naijum Sha supported in part by NSF
Grant CMMI-0654417. Terrance Savitsky was supported under NIH Grant NCI T32 CA096520
while at Rice University.



\end{document}